\documentclass[twocolumn,superscriptaddress,prd,amsmath,amsfonts,nofootinbib]{revtex4-1}

\newcommand{\ud}{\text{d}}

\newcommand{\gb}{\check{g}}
\newcommand{\bs}{\begin{split}}
\newcommand{\es}{\end{split}}
\newcommand{\delb}{\check{\nabla}}

\newcommand{\mc}[1]{\mathcal{#1}}
\newcommand{\bc}[1]{\check{\mathcal{#1}}}
\newcommand{\TT}{\textsc{tt}}
\newcommand{\hb}{{\bar{h}}}

\newcommand{\mub}{{\bar{\mu}}}
\newcommand{\nub}{{\bar{\nu}}}
\newcommand{\alphab}{{\bar{\alpha}}}
\newcommand{\betab}{{\bar{\beta}}}

\usepackage{float}
\usepackage{color}
\usepackage[ocgcolorlinks]{hyperref}
\hypersetup{
    colorlinks,
    citecolor=blue,
    filecolor=blue,
    linkcolor=blue,
    urlcolor=blue,
    pdftitle={Localized Energetics of Linear Gravity: Theoretical Development},
    pdfauthor={Luke M. Butcher}}
\usepackage{slashed}
\usepackage{graphicx}
\usepackage{subfigure}
\usepackage{psfrag}
\usepackage{amsthm}
\theoremstyle{plain}

\begin{document}

\title{Localized Energetics of Linear Gravity: Theoretical Development}
\author{Luke M. Butcher}
\email[]{l.butcher@mrao.cam.ac.uk}
\affiliation{Astrophysics Group, Cavendish Laboratory, J J Thomson Avenue, Cambridge, CB3 0HE, UK}
\affiliation{Kavli Institute for Cosmology, Madingley Road, Cambridge, CB3 0HA, UK}
\author{Michael Hobson}
\affiliation{Astrophysics Group, Cavendish Laboratory, J J Thomson Avenue, Cambridge, CB3 0HE, UK}
\author{Anthony Lasenby}
\affiliation{Astrophysics Group, Cavendish Laboratory, J J Thomson Avenue, Cambridge, CB3 0HE, UK}
\affiliation{Kavli Institute for Cosmology, Madingley Road, Cambridge, CB3 0HA, UK}
\date{\today}

\begin{abstract}
We recently developed a local description of the energy, momentum and angular momentum carried by the linearized gravitational field, wherein the gravitational energy-momentum tensor displays positive energy-density and causal energy-flux, and the gravitational spin-tensor describes purely spatial spin \cite{Butcher10,Ang1}. We now investigate the role these tensors play in a broader theoretical context, demonstrating for the first time that (a) they do indeed constitute Noether currents associated with the symmetry of the linearized gravitational field under translation and rotation, and (b) they are themselves a source of gravity, analogous to the energy-momentum and spin of matter. To prove (a) we construct a Lagrangian for linearized gravity (a covariantized Fierz-Pauli Lagrangian for a massless spin-2 field) and show that our tensors can be obtained from this Lagrangian using a standard variational technique for calculating Noether currents. This approach generates formulae that \emph{uniquely} generalize our gravitational energy-momentum tensor and spin tensor beyond harmonic gauge: we show that no other generalization can be obtained from a covariantized Fierz-Pauli Lagrangian without introducing second derivatives in the energy-momentum tensor. We then construct the Belinfante energy-momentum tensor associated with our framework (combining spin and energy-momentum into a single object) and as our first demonstration of (b) we establish that this Belinfante tensor appears as the second-order contribution to a perturbative expansion of the Einstein field equations, generating the gravitational field in a manner equivalent to the (Belinfante) energy-momentum tensor of matter. By considering a perturbative expansion of the Einstein-Cartan field equations, we then demonstrate that (b) can be realized without forming the Belinfante tensor: our energy-momentum tensor and spin tensor appear as the quadratic terms in \emph{separate} field equations, generating gravity as distinct entities.  Finally, we examine the role of field redefinitions within these perturbative expansions; in contrast to our tensors, the Landau-Lifshitz tensor is found to require a \emph{non-local} field redefinition in order to be cast as a source of the gravitational field. In an appendix, we also give a brief treatment of the global quantities that our framework defines, and verify their equivalence (within the quadratic approximation) to the ADM energy-momentum and angular momentum.
\end{abstract}
\maketitle

\section{Introduction}\label{intro}
There are at least three ways to quantify the energy of a physical system. One approach is to consider interactions with a second system (the energy of which is known) and to seek a function-of-state of the first system which, by undergoing equal and opposite changes, accounts for the energy lost or gained by the second system. Alternatively, a Lagrangian for the physical system might be constructed, and the energy identified as the Noether charge associated with the Lagrangian's symmetry under translation in time. Thirdly, and most simply of all, one can ``weigh'' the system; the energy is then determined by the gravity it generates.

In the preceding articles \cite{Butcher10, Ang1}, we arrived at a local description of the energy, momentum, and angular momentum of the linearized gravitational field.\footnote{To clarify: we have not performed the demonstrably \emph{impossible} feat of finding an energy-momentum tensor $\tau_{ab} \sim \delb h \delb h$ and spin tensor $s^a_{\phantom{a}bc} \sim h\delb h$ that are invariant under the linearized gauge transformations of the gravitational field $\delta h_{ab}=\delb_{(a}\xi_{b)}$. Rather, we rely on a gauge-fixing programme (motivated by key properties of $\tau_{ab}$ and $s^a_{\phantom{a}bc}$, and the energetics of an infinitesimal gravitational detector) to remove the freedom to perform such transformations, and hence arrive at a physically unambiguous description. }  The resulting gravitational energy-momentum tensor $\tau_{ab}$ and spin tensor $s^a_{\phantom{a}bc}$ are particularly notable in that, whenever the field is transverse-traceless, they describe non-negative energy-density, causal energy flow, and spatial spin; moreover, these properties, and the gauge invariant energetics of an infinitesimal probe, motivate a natural gauge-fixing procedure. These two tensors were derived by what is essentially the first method described above: we sought functions of the gravitational field which could account for the energy, momentum, and angular momentum exchanged locally with matter. The main purpose of this present investigation is to demonstrate that $\tau_{ab}$ and $s^a_{\phantom{a}bc}$ fulfil the other two roles of energy-momentum and spin: (a) they are Noether currents of local translations and rotations, and (b) they appear as self-interaction terms in the field equations, generating gravity alongside the energy-momentum and spin of matter.

The paper is organized as follows. In section \ref{Lag}, we construct a Lagrangian for linear gravity (a covariantization of the Fierz-Pauli Lagrangian for a massless spin-2 field) and show that it generates $\tau_{ab}$ and $s^a_{\phantom{a}bc}$ according to standard variational definitions of energy-momentum and spin. This process of ``gauging'' the translational and rotational symmetries of the background, and deriving  $\tau_{ab}$ and $s^a_{\phantom{a}bc}$ as functional derivatives of the Lagrangian with respect to the gauge fields, confirms their status as Noether currents of translational and rotational symmetry. (More precisely, the vectors $\tau_{\mu}{}^a$ are the Noether currents of translations in the $x^\mu$ direction, and $j_{\mu\nu}^{\phantom{\mu\nu}a}\equiv 2x_{[\mu}\tau_{\nu]}{}^a + s^a_{\phantom{a}\mu\nu}$ are the Noether currents of rotations in the $x^\mu x^\nu$-plane.)  In section \ref{Source} we then demonstrate that, under a local redefinition of the gravitational field, $\tau_{ab}$ and $s^{a}_{\phantom{a}bc}$ appear (combined into a single Belinfante energy-momentum tensor) as the \emph{quadratic} part of the vacuum Einstein field equations. The same techniques are then applied to Einstein-Cartan gravity, with $\tau_{ab}$ and $s^{a}_{\phantom{a}bc}$ appearing as the quadratic parts of two \emph{separate} field equations. In both cases, $\tau_{ab}$ and $s^{a}_{\phantom{a}bc}$ quantify the self-interaction of the gravitational field, and generate gravity in an identical fashion to material energy-momentum and spin. Finally, in section \ref{Redef} we examine the significance of the field redefinitions used in section \ref{Source}; in contrast to our tensors, we show that a \emph{non-local} redefinition is required in order to cast the Landau-Lifshitz tensor \cite{LL} as a source of the gravitational field.

From all of this we learn that $\tau_{ab}$ and $s^a_{\phantom{a}bc}$ stand on an equal footing with the energy-momentum and intrinsic spin of matter: they can be derived from the symmetries of a suitable Lagrangian, and behave as sources for the gravitational field. These developments solidify the tensors' physical interpretation, and embed them within the same theoretical  apparatus that has been used to define gravitational energy-momentum in the past \cite{Einstein,LL,DiracPsuedo, ADMII}.  Furthermore, through their role in the non-linear field equations, we gain insight into how $\tau_{ab}$ and $s^{a}_{\phantom{a}bc}$ may be extended beyond the linear regime, and also uncover a new and possibly valuable set of gravitational field-variables.

Our notation and conventions have not changed since the previous papers \cite{Butcher10, Ang1}. We will, however, introduce a new variety of index: Greek letters with overbars, which we will use to enumerate the components of tensors in the non-holonomic basis defined by the tetrad $e_{\mub}^a$; see appendix \ref{ECintro} for details. (These bars should not to be confused with those placed over 2-index tensors, which indicate trace-reversal: $\bar{h}_{ab}\equiv h_{ab} - \gb_{ab} h/2$ and $\bar{\tau}_{ab}\equiv \tau_{ab} - \gb_{ab} \tau/2$.) As before, when working in the flat background spacetime $(\bc{M},\gb_{ab})$ it will often be convenient to express our tensors in a coordinate system $\{x^\mu \}$ that is Lorentzian with respect to the background metric: $\gb_{\mu\nu}= \eta_{\mu\nu}$.

\section{Lagrangian Formulation}\label{Lag}
The primary aim of this section is to reproduce the formulae obtained in \cite{Butcher10, Ang1},
\begin{subequations}\label{tau+s}
\begin{align}\label{tau}
\kappa \bar{\tau}_{\mu\nu}&=\tfrac{1}{4}\partial_\mu  h_{\alpha\beta}\partial_\nu \hb^{\alpha\beta},\\\label{s}
\kappa s^\alpha_{\phantom{\alpha}\mu\nu} &= 2 \hb_{\beta[\nu}\partial^{[\alpha} \hb_{\mu]}{}^{\beta]},
\end{align}
\end{subequations}
from a Lagrangian $\mc{L}$ that also generates the field equations of linear gravity.\footnote{It will often be convenient to express the formula for $\tau_{\mu\nu}$ in terms of its trace-reverse $\bar{\tau}_{\mu\nu}$, as we have in equation (\ref{tau}). Although the algebraically simpler object $\bar{\tau}_{\mu\nu}$ contains all the structure needed to reconstruct the energy-momentum tensor $\tau_{\mu\nu}$, it does not itself carry a physical interpretation.} Before we do this, however, we will have to make some accommodation for the harmonic gauge condition,
\begin{align}\label{harmonic}
\partial^\mu \bar{h}_{\mu\nu}=0,
\end{align}
which we have been diligently enforcing since its appearance as an unexpected consequence of the derivation of $\tau_{\mu\nu}$ \cite{Butcher10}. While this condition has been immensely valuable, naturally reducing the gauge ambiguity of our framework, it now has the potential to interfere with the arbitrary field variations that occur when taking functional derivatives of a Lagrangian. To avoid this problem, we shall temporarily relax the gauge condition (\ref{harmonic}) and aim to derive from $\mc{L}$ an energy-momentum tensor and spin tensor that \emph{generalize} $\tau_{\mu\nu}$ and $s^\alpha_{\phantom{\alpha}\mu\nu}$ beyond the harmonic gauge, reducing to the familiar formulae (\ref{tau+s}) only once the harmonic condition is reintroduced. It is worth noting, however, that although the generalized forms of the tensors will be useful for technical reasons in later sections, they will not give us any further physical information than their restriction to harmonic gauge (\ref{tau+s}). This is because in order to interpret $\tau_{\mu\nu}$ and $s^\alpha_{\phantom{\alpha}\mu\nu}$ physically, we must first extinguish their gauge freedom; the only way to do this that produces sensible local properties (positive energy-density, causal energy-flow, and spatial spin) is by insisting on transverse-traceless gauge, which obviously ensures that the harmonic condition is satisfied.

In addition to relaxing the harmonic condition, it will also be convenient to ignore matter ($T_{\mu\nu}=0$) and work with gravity \emph{in vacuo} for the entirety of this section. Even though the framework of our previous papers was developed around the exchange of energy-momentum and angular momentum between matter and gravity, here we will be able to construct $\tau_{\mu\nu}$ and $s^\alpha_{\phantom{\alpha}\mu\nu}$ from the dynamics of the gravitational field alone.

\subsection{The Fierz-Pauli Lagrangian}\label{FPLagsec}
We begin in a flat background spacetime $(\bc{M},\gb_{ab})$ with the Fierz-Pauli Lagrangian for a massless spin-2 field \cite{fierz}: 
\begin{align}\label{FP}
\mc{L}_\mathrm{FP}\equiv \frac{1}{4\kappa} \left(\partial_\mu h_{\alpha\beta} \partial^\mu \hb^{\alpha\beta} - 2 \partial_\mu \hb^{\mu\alpha}\partial_\nu \hb^\nu_{\phantom{\nu}\alpha}\right).
\end{align}
From a non-gravitational standpoint, this Lagrangian can be derived 
by demanding invariance under the massless spin-2 gauge transformation \cite{Pad}:
\begin{align}\label{spin2GT}
\delta h_{\mu\nu}= \partial_{(\mu} \xi_{\nu)} \quad \Rightarrow \quad \delta \mc{L}_\mathrm{FP} = \text{surface terms}.
\end{align}
For our purposes, however, it suffices to observe that $\mc{L}_\mathrm{FP}$ correctly reproduces the linearized vacuum Einstein field equations:
\begin{align}\label{LinFE}
0=\frac{\delta\mc{L}_\mathrm{FP}}{\delta h_{\mu\nu}}=\frac{1}{\kappa} \widehat{G}^{\mu\nu\alpha\beta}h_{\alpha\beta},
\end{align}
where 
\begin{align}\nonumber
\widehat{G}_{\mu\nu}^{\phantom{\mu\nu}\alpha\beta}h_{\alpha\beta} \equiv G^{(1)}_{\mu\nu} &\equiv\partial_\alpha \partial_{(\mu}h_{\nu)}{}^{\alpha} - \partial^2 h_{\mu\nu}/2 - \partial_\mu\partial_\nu h/2
\\\label{Gdef}&\quad+ \eta_{\mu\nu}(\partial^2 h - \partial_\alpha\partial_\beta h^{\alpha\beta} )/2
\end{align}
is the linear part of the Einstein tensor when the physical metric $g_{ab}$ is perturbed according to
\begin{align}\label{hdef}
\phi^* g_{ab}=\gb_{ab} + h_{ab}.
\end{align}
As before, $\phi: \mc{M}\to \bc{M}$ maps the physical spacetime $(\mc{M},g_{ab})$ to the background.

We wish to obtain an energy-momentum tensor and a spin tensor from $\mc{L}_\mathrm{FP}$ by ``gauging'' the symmetries of the background spacetime, and taking functional derivatives with respect to the background gauge fields. In the standard formulation of general relativity, the rotational and translational symmetries of Minkowski spacetime are gauged into a \emph{single} local symmetry: \emph{diffeomorphism gauge invariance}.  As a result, the energy-momentum and angular momentum of matter are all contained within a single \emph{Belinfante} energy-momentum tensor $T^\mathrm{Bel}_{ab}$ (written simply as $T_{ab}$ in \cite{Butcher10,Ang1}) which one derives according to Hilbert's definition: $T^\mathrm{Bel}_{ab}\equiv (1/\sqrt{-g})(\delta \mc{L}_\mathrm{matter}/ \delta g^{ab})$.  In order to obtain two separate tensors, $\tau_{\mu\nu}$ and $s^\alpha_{\phantom{\alpha}\mu\nu}$, following a Hilbert-like approach, one must gauge the rotational and translational symmetries \emph{separately}, and take derivatives with respect to the two gauge fields (\ref{T+Sdef}). These are the techniques of  Einstein-Cartan gravity (as formulated by Kibble and Sciama \cite{Kibble, Sciama1,Sciama2, HehlEC, BlagEC, GTG, GTG2}) which we summarize in appendix \ref{ECintro}.

To apply these techniques to the problem at hand, we will first need to covariantize $\mc{L}_\mathrm{FP}$. Invoking a background tetrad $\check{e}^a_\mub$ and spin connection $\check{\omega}_a^{\phantom{a}\mub\nub}$, we write the Fierz-Pauli Lagrangian in terms of quantities which are covariant under local translations and rotations:
\begin{align}\label{FPcov}
\mc{L}^\prime_\mathrm{FP}&\equiv \frac{\check{e}}{4\kappa} \left(\check{D}_\mub h_{\alphab\betab} \check{D}^\mub \hb^{\alphab\betab} - 2 \check{D}_\mub \hb^{\mub\alphab}\check{D}_\nub \hb^\nub_{\phantom{\nub}\alphab}\right),
\end{align}
where $\check{D}_a$ is a covariant derivative with connection $\check{\omega}_a^{\phantom{a}\mub\nub}$, the volume element $\check{e}\equiv 1/\det(\check{e}^a_\mub)$, and Greek indices with overbars enumerate the components of tensors in the non-holonomic basis $\{\check{e}^a_\mub \}$. 

In order to perform arbitrary infinitesimal variations in $\check{e}^a_\mub$ and $\check{\omega}_a^{\phantom{a}\mub\nub}$, we will inevitably explore backgrounds with non-zero curvature $\check{R}_{ab\mub\nub}$ and torsion $\check{\mc{T}}^a{}_{\mub\nub}$. For this reason, we must also decide how our Lagrangian should change when the background is no longer flat and torsion-free.\footnote{One might hope to avoid this decision by instead demanding that the variations in $\check{e}^a_{\mub}$ and $\check{\omega}_a^{\phantom{a}\mub\nub}$ be constrained to backgrounds with $\check{R}_{ab\mub\nub}=\check{\mc{T}}^a_{\phantom{a}\mu\nu}=0$. However, one must enforce this restriction by including Lagrange multiplier terms $\lambda^{ab\mub\nub}\check{R}_{ab\mub\nub} + \lambda_a^{\phantom{a}\mub\nub}\check{\mc{T}}^a_{\phantom{a}\mu\nu}$ in the Lagrangian, which inevitably produce unconstrained \emph{superpotentials} proportional to $\partial_\beta \lambda_{\mu\nu}{}^{\beta}$ in the energy-momentum tensor and  $\partial_\beta \lambda^{\alpha\beta}{}_{\mu\nu} - \lambda_{[\mu\nu]}{}^\alpha$  in the spin tensor. Thus, one cannot use these ``constrained variations'' alone to construct $\tau_{\mu\nu}$ and $s^\alpha_{\phantom{\alpha}\mu\nu}$ from a Lagrangian, as one must make further specifications of the form $\lambda=\ldots$ before the energy-momentum tensor and spin tensor are well-defined. In contrast, ``unconstrained variations'' \emph{can} be used to construct $\tau_{\mu\nu}$ and $s^\alpha_{\phantom{\alpha}\mu\nu}$, as we will soon see. The key to this approach is that we are forced to describe how $\mc{L}^\prime_{\mathrm{FP}}$ behaves when the background is curved and contorted. Because there is a precise correspondence between background-coupling and superpotentials (see section \ref{supers}) once this behaviour of the Lagrangian is fixed, it is able to define the energy-momentum tensor and spin tensor unambiguously.} The obvious response to this uncertainty is to follow the ``minimal coupling'' maxim, and insist that the Lagrangian remain as it is in equation (\ref{FPcov}) even when the background is curved and contorted. Despite the simplicity of this approach, the Lagrangian $\mc{L}^\prime_\mathrm{FP}$ is actually a highly unnatural choice, as it deprives the field theory of its spin-2 gauge-invariance when the background is no longer flat. To see this, consider a curved vacuum background ($\check{R}_{a\mub}=0$, $\check{\mc{T}}^a{}_{\mub\nub}=0$, $\check{R}_{ab\mub\nub}\ne0$) and perform a covariantized spin-2 gauge transformation:
\begin{align}\nonumber
\delta h_{\mub\nub}= \check{D}_{(\mub} \xi_{\nub)} \quad \Rightarrow\quad \delta \mc{L}^\prime_\mathrm{FP}&= -\frac{\check{e}}{\kappa} \check{D}^{\mub} \xi^{\nub} \check{R}_{\alphab\mub\nub\betab} \hb^{\alphab\betab}\\\label{GIfail}&\quad {} + \text{surface terms.}
\end{align}
Thus, $ \mc{L}^\prime_\mathrm{FP}$ loses its spin-2 gauge invariance when one tries to extend the theory ``minimally'' beyond the flat background.

The gauge invariance of the field theory can be preserved, for vacuum backgrounds at least, if we allow $h_{\mu\nu}$ to couple directly to the curvature of the background.\footnote{No set of curvature-coupling terms (nor torsion-coupling terms) can extend the theory's gauge invariance to include \emph{non-vacuum} backgrounds. This comes as no surprise, considering that we are studying a Lagrangian $\mc{L}_\mathrm{FP}$ that does not include matter. Evidently, the linearized vacuum field equations (\ref{LinFE}) can only be expected to be consistent when they describe perturbations from a vacuum background.} The Lagrangian (\ref{Ldef}) we will use to generate $\tau_{\mu\nu}$ and $s^\alpha_{\phantom{\alpha}\mu\nu}$ will do exactly this, although we should explain that it is not unique in this regard. If we had wanted to present the subsequent calculation as a genuine \emph{ab initio} derivation of $\tau_{\mu\nu}$ and $s^\alpha_{\phantom{\alpha}\mu\nu}$, then we would need to justify our specialisation to (\ref{Ldef}) over the other possibilities. However, we have already derived $\tau_{\mu\nu}$ and $s^\alpha_{\phantom{\alpha}\mu\nu}$ from more concrete considerations \cite{Butcher10, Ang1}, and our aim here is only to show that a Lagrangian \emph{exists} from which the tensors can be obtained. We will explore this curvature-coupling freedom in section \ref{supers}, and by the end of section \ref{EHexp} we will be in a position to look back at $\mc{L}$ and better understand the significance of our ``choice''. For now, we shall simply write down our Lagrangian as an ansatz, justified by its being a covariantization of $\mc{L}_\mathrm{FP}$ which preserves the field theory's gauge-invariance beyond the flat background, and proceed to calculate its energy-momentum tensor and spin tensor.

\subsection{Energy-Momentum Tensor and Spin Tensor}
Let us consider the following the Lagrangian for the linearized gravitational field:
\begin{align}\nonumber
\mc{L}&\equiv \frac{\check{e}}{4\kappa} \left(\check{D}_\mub h_{\alphab\betab} \check{D}^\mub \hb^{\alphab\betab} - 2 \check{D}_\mub \hb^{\mub\alphab}\check{D}_\nub \hb^\nub_{\phantom{\nub}\alphab}\right.\\\label{Ldef}&\qquad\quad \left. {} + 2 \hb^{\mub\nub} \check{R}_{\alphab\mub\nub\betab} \hb^{\alphab\betab} \right).
\end{align}
This clearly reduces to the Fierz-Pauli Lagrangian (\ref{FP}) when the background is flat and torsion-free, and furthermore, successfully extends the spin-2 gauge-invariance of the theory to curved (vacuum) backgrounds:
\begin{align}\label{GIsucc}
\delta h_{\mub\nub}= \check{D}_{(\mub} \xi_{\nub)} \quad \Rightarrow\quad \delta \mc{L}= \text{surface terms.}
\end{align}

Treating the fields $\{h^{\mub\nub}, \check{e}^a_\mub, \check{\omega}_a^{\phantom{a}\mub\nub}\}$ as independent variables, we shall evaluate the energy-momentum tensor and spin tensor of $\mc{L}$  according to their definitions from Einstein-Cartan gravity:
\begin{align}\label{tau+sdef}
\tau_a^{\phantom{a}\mu}&\equiv\left(\frac{1}{2 \check{e}}\frac{\delta \mc{L}}{\delta \check{e}_\mub^a}\right)_{\substack{\check{e}=\delta\\\check{\omega}=0}},& s^a_{\phantom{a}\mu\nu}&\equiv \left(
\frac{1}{\check{e}}\frac{\delta \mc{L}}{\delta \check{\omega}_a^{\phantom{a}\mub\nub}}\right)_{\substack{\check{e}=\delta\\\check{\omega}=0}},
\end{align}
where the subscripts $\check{e}=\delta$ and $\check{\omega}=0$ signify that, once the functional derivatives have been taken, we evaluate the tensors on a flat torsion-free background, and the tetrad and spin connection become trivial (\ref{flate}) to reflect this. Substituting (\ref{Ldef}) into (\ref{tau+sdef}) we arrive at the following formulae for the energy-momentum tensor and spin tensor of the linearized gravitational field:
\begin{subequations}\label{newtensors}
\begin{align}\label{gentau}
\kappa \bar{\tau}_{\mu\nu} &= \tfrac{1}{4} \partial_\mu h_{\alpha\beta} \partial_\nu \hb^{\alpha\beta} - \tfrac{1}{2} \partial_\mu \hb_{\nu\alpha} \partial_\beta \hb^{\alpha\beta},\\\label{gens}
\kappa s^\alpha_{\phantom{\alpha} \mu\nu}&= 2\hb_{\beta[\nu} \partial^{[\alpha} \hb_{\mu]}{}^{\beta]} + \delta_{[\nu}{\!\!\!}^\alpha\hb_{\mu]}{}^\beta\partial_\gamma \hb^\gamma_{\phantom{\gamma}\beta}.
\end{align}
\end{subequations}
This is precisely the result we needed: $\mc{L}$ has generated an energy-momentum tensor $\tau_{\mu\nu}$ and spin tensor $s^\alpha_{\phantom{\alpha}\mu\nu}$ which reduce to the familiar formulae (\ref{tau+s}) when the harmonic condition (\ref{harmonic}) is reintroduced.

We have achieved the main aim of this section, demonstrating that our energy-momentum tensor and spin tensor can be identified as translational and rotational Noether currents of a Lagrangian for linear gravity.  In addition, the equations (\ref{newtensors}) reveal how our tensors (\ref{tau+s}) generalize beyond harmonic gauge. Before we study these generalized tensors in detail, we shall first examine the freedom that was available in our choice of covariantization of $\mc{L}_\mathrm{FP}$, and demonstrate that the formulae (\ref{newtensors}) constitute a suitably unique extension of (\ref{tau+s}).

\subsection{Background Coupling and Superpotentials}\label{supers}
We begin by considering the most general Lagrangian, quadratic in $h_{\mu\nu}$ and second-order in derivatives, which differs from the minimally coupled Lagrangian (\ref{FPcov}) only by terms which couple $h_{\mu\nu}$ to background curvature; ignoring surface terms, this is
\begin{align}\label{LRdef}
\mc{L}_{\check{R}} \equiv \mc{L}^\prime_\mathrm{FP} +\frac{\check{e}}{2\kappa}\check{R}_{\alphab\betab}^{\phantom{\alpha\beta}
\mub\nub} \Sigma^{\alphab\betab}{}_{\mub\nub},
\end{align} 
where $\Sigma^{\alpha\beta}{}_{\mu\nu}=-\Sigma^{\beta\alpha}{}_{\mu\nu}=-\Sigma^{\alpha\beta}{}_{\nu\mu}=\Sigma_{\mu\nu}{}^{\beta\alpha}$ is a local quadratic Lorentz-covariant function of $h_{\mu\nu}$, the general form of which can be parametrized by five dimensionless constants $\{A_n\}$:
\begin{align}\nonumber
\Sigma^{\alpha\beta}{}_{\mu\nu}&\equiv A_1 h^\alpha_{[\mu}h_{\nu]}^\beta + A_2 h h^{[\alpha}_{[\mu}\delta^{\beta]}_{\nu]}+A_3 h^\gamma_{[\mu}\delta_{\nu]}^{[\beta}h^{\alpha]}_\gamma \\\label{Sigma}
&\quad {} + \delta^{\alpha}_{[\mu}\delta^{\beta}_{\nu]}\left(A_4 h^2 + A_5 h_{\gamma\delta}h^{\gamma\delta}\right).
\end{align}
If we recall the behaviour of $\mc{L}^\prime_\mathrm{FP}$ under a spin-2 gauge transformation (\ref{GIfail}), it is immediately clear that the field theory will retain its gauge invariance for curved vacuum backgrounds (\ref{GIsucc}) if and only if $A_1=-1$.

Inserting $\mc{L}_{\check{R}}$ into (\ref{tau+sdef}), we find that the energy-momentum tensor of this Lagrangian is identical to the tensor (\ref{gentau}) derived from $\mc{L}$, but that the spin tensor is given by
\begin{align}\nonumber
\kappa s^\alpha_{\phantom{\alpha} \mu\nu}&= h_{\beta[\nu}\partial^\alpha h_{\mu]}{}^\beta  + \delta_{[\nu}{\!\!\!}^\alpha\hb_{\mu]}{}^\beta\partial_\gamma \hb^\gamma_{\phantom{\gamma}\beta}+ \hb^\alpha{}_{[\mu}\partial^\beta \hb_{\nu]\beta}\\\label{gens2}& \quad {} + \partial_\beta \Sigma^{\alpha\beta}{}_{\mu\nu}.
\end{align}
In harmonic gauge this becomes
\begin{align}\label{gens3}
\kappa s^\alpha_{\phantom{\alpha}\mu\nu} = h_{\beta[\nu}\partial^\alpha h_{\mu]}{}^\beta + \partial_\beta \Sigma^{\alpha\beta}{}_{\mu\nu},
\end{align}
revealing that $\{A_n\}$ are the very same constants that parameterized the \emph{superpotential} freedom of $s^\alpha_{\phantom{\alpha}\mu\nu}$ in section III of our previous paper \cite{Ang1}. There, the value $A_1=-1$ was fixed by demanding that $s^\alpha_{\phantom{\alpha}\alpha\nu}=0$ for all transverse-traceless $h_{\mu\nu}$. Now we see that this special value of $A_1$ has yet another significance: it ensures that the spin-2 gauge-invariance of the linearized theory extends beyond the flat background. 

Equation (\ref{gens3}) also demonstrates that the parameters $\{A_n\}$ must take the values
\begin{align}\label{An=}
A_1=-1,\ \ A_2=1, \ \ A_4= -1 /4, \ \ A_3=A_5=0,
\end{align}
(as they do in \cite{Ang1}) if the spin tensor (\ref{gens2}) is to reduce to its original form (\ref{s}) in harmonic gauge; thus the freedom to add curvature terms (\ref{LRdef}) cannot, by itself, produce any other generalization of $\tau_{\mu\nu}$ and $s^\alpha_{\phantom{\alpha}\mu\nu}$ than (\ref{newtensors}).

Now that we understand the role played by curvature terms in the Lagrangian, we must also explore the possibility of coupling $h_{\mu\nu}$ to background torsion. If the Lagrangian is to remain quadratic in $h_{\mu\nu}$ and second-order in derivatives, the only contribution we need to consider is
\begin{align}\label{torsionterms}
\Delta \mc{L} \equiv  - \frac{\check{e}}{\kappa} \bc{T}^a_{\phantom{a}\mub\nub} \Sigma_a^{\phantom{a}\mub\nub},
\end{align}
where $\Sigma_a^{\phantom{a}\mu\nu}= - \Sigma_a^{\phantom{a}\nu\mu}$ is composed of terms of the form $hDh$. The torsion terms generate the superpotential freedom of the energy-momentum tensor:
\begin{align}\label{supertau}
\kappa \Delta \tau_a^{\phantom{a}\mu}= \left(\frac{\kappa}{2 \check{e}}\frac{\delta \Delta \mc{L}}{\delta \check{e}_\mub^a}\right)_{\substack{\check{e}=\delta\\\check{\omega}=0}} = \partial_\nu \Sigma_a{}^{\mu\nu},
\end{align} 
which is also accompanied by a change in the spin tensor,
\begin{align}\label{supertaus}
\kappa \Delta s^a_{\phantom{a}\mu\nu}&= \left(
\frac{\kappa}{\check{e}}\frac{\delta\Delta \mc{L}}{\delta \check{\omega}_a^{\phantom{a}\mub\nub}}\right)_{\substack{\check{e}=\delta\\\check{\omega}=0}} = 2\Sigma_{[\mu\nu]}{}^a.
\end{align}
Because the energy-momentum superpotentials are of the form $\partial (h\partial h)$, containing tensors of the form $h \partial^2 h$, their addition has the potential to spoil the homogeneous differential structure of (\ref{gentau}): $\tau_{\mu\nu} \sim \partial h\partial h$. In fact, there is no superpotential $\partial_\nu \Sigma_a{}^{\mu\nu}$, entirely composed of terms $\partial h\partial h$, which vanishes in harmonic gauge. [To prove this, construct the most general 2-index Lorentz-covariant tensor, composed entirely of terms of the form $\partial h \partial h$, which is at least linear in $\partial_\mu \hb^{\mu\nu}$, and suppose that it is also a superpotential: $\kappa \Delta \tau_{\mu\nu}= \partial_\alpha \hb^{\alpha\beta} (C_1 \partial_\mu \hb_{\nu \beta} + C_2\partial_\nu \hb_{\mu\beta} + C_3 \partial_\beta \hb_{\mu\nu} + \eta_{\mu\nu}(C_4 \partial_\beta h + C_5 \partial_\gamma \hb^\gamma_{\phantom{\gamma} \beta})) + C_6 \partial_\alpha \hb^\alpha_{\phantom{\alpha} \mu} \partial_\nu h + C_7 \partial_\alpha \hb^\alpha_{\phantom{\alpha} \nu}\partial_\mu h + C_8 \partial_\alpha\hb^\alpha_{\phantom{\alpha} \mu}\partial_\beta\hb^\beta_{\phantom{\beta} \nu}$, where $\{C_n\}$ are arbitrary dimensionless constants. Equation (\ref{supertau}) informs us that $\partial_\nu \Delta \tau_{\mu}^{\phantom{\mu}\nu}=0$ for all $h_{\mu\nu}$; the only values of $\{C_n\}$ consistent with this are $C_n=0$.] Thus, the freedom to add torsion terms to the Lagrangian is nullified by our insistence that the generalized $\tau_{\mu\nu}$ be free of second derivatives, and reduce to our original formula when the harmonic condition is enforced. 

We therefore conclude that our Lagrangian (\ref{Ldef}) is the \emph{unique} covariantization of $\mc{L}_\mathrm{FP}$, quadratic in $h_{\mu\nu}$ and second-order in derivatives, which according to the definitions (\ref{tau+sdef}) generates an energy-momentum tensor that is free from second derivatives, and an energy-momentum tensor and spin tensor which agree with our original formulae (\ref{tau+s}) in harmonic gauge. Consequently, the resulting energy-momentum tensor and spin tensor (\ref{newtensors}) are the unique extensions of $\tau_{\mu\nu}$ and $s^\alpha_{\phantom{\alpha}\mu\nu}$ beyond harmonic gauge, which can be derived from a covariantized Fierz-Pauli Lagrangian according (\ref{tau+sdef}), and which do not introduce terms of the form $h\partial^2 h$ into $\tau_{\mu\nu}$.

Having demonstrated that (\ref{newtensors}) are indeed the unique extension of (\ref{tau+s}) beyond the harmonic gauge, it will be useful to construct their Belinfante tensor. 

\subsection{Belinfante Tensor}\label{Bel}
As an alternative to our description of gravitational energy-momentum and spin in terms of two separate tensors, $\tau_{\mu\nu}$ and  $s^\alpha_{\phantom{\alpha}\mu\nu}$, we may construct a \emph{Belinfante} energy-momentum tensor \cite{belinfante},
\begin{align}\label{Beldef}
t_{\mu\nu}\equiv \tau_{\mu\nu} + \partial_\alpha (s_{\mu\nu}^{\phantom{\mu\nu}\alpha} + s_{\nu\mu}^{\phantom{\mu\nu}\alpha} - s^\alpha_{\phantom{\alpha}\mu\nu})/2,
\end{align}
which combines the two. This tensor is symmetric by virtue of the field equations,
\begin{align}\nonumber
t_{[\mu\nu]}&= \tau_{[\mu\nu]} - \partial_\alpha s^\alpha_{\phantom{\alpha}\mu\nu}/2\\ &= \frac{1}{\kappa} h_{\alpha[\nu} \widehat{G}_{\mu]}{}^{\alpha\beta\gamma}h_{\beta\gamma} =0,
\end{align}
and is also conserved:
\begin{align}
\partial^\nu t_{\mu\nu} &= \partial^\nu \tau_{\mu\nu} = -\frac{1}{2\kappa} (\partial_\mu h^{\alpha\beta})\widehat{G}_{\alpha\beta}^{\phantom{\alpha\beta}\gamma\delta}h_{\gamma\delta} =0.
\end{align}
Furthermore, provided surface terms are negligible, the Belinfante tensor defines precisely the same \emph{global} measure of energy, momentum, and angular momentum as $\tau_{\mu\nu}$ and $s^\alpha_{\phantom{\alpha}\mu\nu}$:
\begin{subequations}\label{global}
\begin{align}\label{globalEM}
\int t_{\mu}^{\phantom{\mu}0}\ud^3x &= \int \tau_{\mu}^{\phantom{\mu}0} \ud^3x,\\\label{globalAM}
\int 2 x_{[\mu} t_{\nu]}{}^0\ud^3x &= \int( 2x_{[\mu}\tau_{\nu]}{}^0 + s^0_{\phantom{0}\mu\nu}) \ud^3x.
\end{align}
\end{subequations}

The advantage of this Belinfante description is obvious: it combines energy-momentum and spin into a single symmetric tensor.\footnote{That said, as far as our framework is concerned, the symmetry of the Belinfante tensor is not particularly impressive: our energy-momentum tensor $\tau_{\mu\nu}$ is already symmetric, by virtue of the harmonic condition (\ref{harmonic}) rather than the field equations.} This apparent simplicity comes at a high price, however, because although the global picture remains intact (\ref{global}) the Belinfante tensor is unable to reproduce the physically sensible \emph{local} description that $\tau_{\mu\nu}$ and $s^\alpha_{\phantom{\alpha}\mu\nu}$ provide.

In general, the intermixture of spin and energy-momentum in (\ref{Beldef}) prevents us from localizing the two quantities separately, and we are left with angular momentum currents $x_{[\mu} t_{\nu]}{}^\alpha$ which ``contain'' spin but do not assign it a local current, and energy-momentum currents $t_\mu^{\phantom{\mu}\alpha}$ which display \emph{negative} energy-densities and \emph{non-causal} energy-flux. Furthermore, because the gravitational Belinfante tensor has no special geometric or algebraic properties in either harmonic or transverse-traceless gauge, it becomes impossible to justify a natural gauge-fixing program. The tensor $t_{\mu\nu}$ can then be evaluated over the entire gauge space of $h_{\mu\nu}$, and will depend on the arbitrary mapping $\phi: \mc{M} \to \bc{M}$ as much as it depends on the physical properties of the gravitational field.

For these reasons, we cannot advocate interpreting $t_{\mu\nu}$ as the ``true'' energy-momentum of the gravitational field. The tensors $\tau_{\mu\nu}$ and $s^\alpha_{\phantom{\alpha}\mu\nu}$ are the local measures of gravitational energy-momentum and spin, describing positive energy-density, causal energy-flux, and spatial spin; $t_{\mu\nu}$ is a derived quantity which packages spin and energy-momentum into a single object, losing some local information in the process. The main application of the gravitational Belinfante tensor will arise in the next section, where we will also gain some insight into its physical interpretation. In brief, we will see that $t_{\mu\nu}$ appears as the quadratic contribution to the Einstein field equations, generating perturbations in the metric alongside the (Belinfante) energy-momentum of matter. In other words, it is the particular combination of energy-momentum and spin, $\tau_{\mu\nu} + \partial_\alpha (s_{\mu\nu}^{\phantom{\mu\nu}\alpha} + s_{\nu\mu}^{\phantom{\nu\mu}\alpha}- s^\alpha_{\phantom{\alpha}\mu\nu})/2$, that curves physical spacetime in a quadratic approximation to general relativity. It would be implausible to expect $\tau_{\mu\nu}$ alone to fulfill this role, as there is no other field equation in which $s^\alpha_{\phantom{\alpha} \mu\nu}$ could act as the source; only by considering perturbations in the Einstein-Cartan equations, as we do in section \ref{ECsource}, will we find a setting in which $\tau_{\mu\nu}$ and $s^\alpha_{\phantom{\alpha}\mu\nu}$ arise as the self-interaction source-terms in two separate field equations. 

Proceeding with the calculation, we substitute the generalized tensors (\ref{newtensors}) into the definition (\ref{Beldef}) in order to obtain the Belinfante tensor associated with our framework. The resulting formula can be expressed most compactly in terms of the trace-reverse of $t_{\mu\nu}$:
\begin{align}\nonumber
\kappa \bar{t}_{\mu\nu}&= \frac{1}{4} \partial_\mu h_{\alpha\beta}\partial_\nu \hb^{\alpha\beta}  + \frac{1}{2}\partial_\alpha \hb^{\alpha\beta}\left( \partial_{(\mu} \hb_{\nu)\beta} 
 - \partial_\beta h_{\mu\nu}\right)
\\\nonumber 
&\quad {}  + \frac{1}{2} \partial_{\alpha}\hb_{\beta(\mu}\left(\partial^\beta \hb_{\nu)}{}^\alpha  - \partial_{\nu)} \hb^{\alpha\beta}\right)\\\nonumber 
&\quad {}
 + \frac{1}{2} \hb^{\alpha\beta}\left( \partial_\alpha \partial_{(\mu} \hb_{\nu)\beta}  -  \partial_\alpha \partial_\beta h_{\mu\nu}\right)
\\\label{t}
&\quad {}  + \frac{1}{2} \hb_{\beta (\mu} \partial^\beta \partial^\alpha \hb_{\nu)\alpha} + h_{\alpha[\nu} \widehat{G}_{\mu]}{}^{\alpha\beta\gamma}h_{\beta\gamma}.
\end{align}
Although the last term can be removed by applying the field equations, we will retain it for the sake of generality. 

This concludes our analysis of the Lagrangian formulation of $\tau_{\mu\nu}$ and $s^\alpha_{\phantom{\alpha}\mu\nu}$. Armed with the results of this section, we are now in a position to ``weigh'' the gravitational field, and investigate the role our tensors play in the non-linear field equations.

\section{Self-interaction in the Gravitational Field Equations}\label{Source}
Here we examine how $\tau_{\mu\nu}$ and $s^\alpha_{\phantom{\alpha}\mu\nu}$ (in their generalized form (\ref{newtensors})) occur as the quadratic terms in a perturbative expansion of the Einstein field equations and also the Einstein-Cartan field equations, generating gravity in exactly the same fashion as material energy-momentum and spin.

As we move from a linear theory of gravity to a quadratic one, it will become important to fix the definition of $h_{\mu\nu}$ more precisely. Until this point, $h_{\mu\nu}$ has been used to signify a perturbation in the physical metric:
\begin{align}\label{gstandard}
\phi^* g_{ab} = \gb_{ab} + h_{ab}.
\end{align}
However, because our framework is based exclusively on gravity in the linear approximation, we could have defined $h_{\mu\nu}$ such that
\begin{align}\label{redef}
\phi^* g_{ab} = \gb_{ab} + h_{ab} + O(h^2),
\end{align}
and arrived at the very same results. For instance, suppose we had decided to work with the field $h^\prime_{\mu\nu}$ that defines a (negative) perturbation in the \emph{inverse} metric:
\begin{align}\label{ginverse0}
\phi^* g^{ab} = \gb^{ab} - h^{\prime ab}.
\end{align}
The equation for the metric would then have been
\begin{align}\label{ginverse}
\phi^* g_{ab} = \gb_{ab} + h^\prime_{ab} + h^\prime_{ac}h^{\prime c}_{\phantom{\prime c}b} +O(h^3),
\end{align}
instead of (\ref{gstandard}), but because $h^\prime_{\mu\nu}= h_{\mu\nu} + O(h^2)$, the linearized theory of $h^\prime_{\mu\nu}$ would be the same as $h_{\mu\nu}$, and our framework would assign the same tensors (\ref{newtensors}) to describe its energy-momentum and spin. Only once we came to study the quadratic approximation of the field equations, as we do now, would any mathematical difference between the fields $h_{\mu\nu}$ and $h^\prime_{\mu\nu}$ have been observed.

For the sake of concreteness, we will start with the standard definition of the gravitational field (\ref{gstandard}) and consider this to be true to all orders of approximation. As we will soon see, however, it is precisely the freedom to make field redefinitions of the form (\ref{redef}) that will allow us to cast $\tau_{\mu\nu}$ and $s^\alpha_{\phantom{\alpha}\mu\nu}$ as the sources of the gravitational field; by the end of the next section, we will have uncovered a new definition of $h_{\mu\nu}$, valid to quadratic order, that is  specially selected by our local description of gravitational energetics. We will examine the wider significance of this definition, and the effects of field redefinition in general, in section \ref{Redef}.

\subsection{The Einstein Equations}\label{Esource}
Consider the vacuum Einstein field equations (in the physical spacetime) expressed in terms of the ``mixed'' Einstein tensor density:
\begin{align}\label{MixedFE}
\sqrt{-g}G_a^{\phantom{a}b} =0.
\end{align}
Mapping this equation to the background, we apply (\ref{gstandard}) to every instance of $\phi^*g_{ab}$, and expand the result in powers of $h_{\mu\nu}$:
\begin{align}\label{vac}
 \widehat{G}_{\mu\nu}^{\phantom{\mu\nu}\alpha\beta}h_{\alpha\beta} + \widetilde{G}^{(2)}_{\mu\nu}+ O(h^3)=0,
\end{align}
where
\begin{align}
\widetilde{G}^{(2)}{\!}_a^{\phantom{a}b}&\equiv\left[\phi^* (\sqrt{-g} G_a^{\phantom{a}b})\right]^{(2)}
\end{align}
is the quadratic part of the mixed Einstein tensor density.

Let us now redefine the gravitational field $h_{\mu\nu}$ by making the replacement
\begin{align}\label{hredef}
h_{\mu\nu}\to h_{\mu\nu} +  h_{\mu\alpha}h^{\alpha}_{\phantom{\alpha}\nu}/2;
\end{align}
this gives rise to a corresponding change in the definition of the metric, 
\begin{align}\label{gnew}
\phi^* g_{ab} = \gb_{ab} + h_{ab} + h_{ac}h^{ c}_{\phantom{ c}b}/2,
\end{align}
and causes the vacuum equations (\ref{vac}) to become
\begin{align}\label{vac2}
\widehat{G}_{\mu\nu}^{\phantom{\mu\nu}\alpha\beta}h_{\alpha\beta} +\widehat{G}_{\mu\nu}^{\phantom{\mu\nu}\alpha\beta}(h_{\alpha\gamma}h^\gamma_{\phantom{\gamma}\beta})/2 + \widetilde{G}^{(2)}_{\mu\nu}=0,
\end{align}
when working to second order.\footnote{To clarify: the tensor $\widetilde{G}^{(2)}_{\mu\nu}$ is still the quadratic part of $\phi^* (\sqrt{-g} G_a^{\phantom{a}b})$ when the metric is expanded according to (\ref{gstandard}); this tensor has exactly the same formula in terms of the new $h_{\mu\nu}$ as it did the old because the replacement (\ref{hredef}) only alters $\widetilde{G}^{(2)}_{\mu\nu}$ by quantities $O(h^3)$, which we neglect.} Moving all quadratic terms to the right-hand side, and making use of the following identity,
\begin{align}\label{Gtid}
 -\widetilde{G}^{(2)}_{\mu\nu} -   \widehat{G}_{\mu\nu}^{\phantom{\mu\nu}\alpha\beta}(h_{\alpha\gamma}h^\gamma_{\phantom{\gamma}\beta})/2= \kappa t_{\mu\nu},
\end{align}
which is derived in appendix \ref{Id}, we find that the quadratic vacuum field equations (\ref{vac2}) are equivalent to
\begin{align}\label{Gh=t}
\widehat{G}_{\mu\nu}^{\phantom{\mu\nu}\alpha\beta}h_{\alpha\beta} = \kappa t_{\mu\nu},
\end{align}
where $t_{\mu\nu}$ is the Belinfante tensor of the gravitational field (\ref{t}) constructed from $\tau_{\mu\nu}$ and $s^\alpha_{\phantom{\alpha}\mu\nu}$.

Equation (\ref{Gh=t}) is exactly what we had hoped to find: gravitational energy-momentum generates gravity in exactly the same fashion as material energy-momentum. To make this comparison transparent, we remind ourselves of the \emph{non-vacuum} field equations at linear order:
\begin{align}\label{nonvac}
\widehat{G}_{\mu\nu}^{\phantom{\mu\nu}\alpha\beta}h_{\alpha\beta} = \kappa T^\mathrm{Bel}_{\mu\nu},
\end{align}
where $T^\mathrm{Bel}_{\mu\nu}$ (written simply as $T_{\mu\nu}$ in \cite{Butcher10,Ang1}) is the Belinfante energy-momentum tensor of matter, mapped to the background. Because $T^\mathrm{Bel}_{\mu\nu}$ is Belinfante, any intrinsic spin carried by matter must be packaged inside this tensor according to the same formula (\ref{Beldef}) that defines the Belinfante tensor of the gravitational field, making the analogy with $t_{\mu\nu}$ extremely close. Furthermore, if one assumes that $T^\mathrm{Bel}_{\mu\nu}$ is of the same order of magnitude as $t_{\mu\nu}\sim O(h^2)$, then at quadratic order the non-vacuum version of (\ref{Gh=t}) is in fact
\begin{align}\label{Gh=t+T}
\widehat{G}_{\mu\nu}^{\phantom{\mu\nu}\alpha\beta}h_{\alpha\beta} = \kappa (t_{\mu\nu} +T^\mathrm{Bel}_{\mu\nu}),
\end{align}
wherein the source of the gravitational field is the sum of the material and gravitational Belinfante tensors.

It goes without saying that equation (\ref{Gh=t}) can also be written as
\begin{align}\label{Gh=tau+s}
\!\widehat{G}_{\mu\nu}^{\phantom{\mu\nu}\alpha\beta}h_{\alpha\beta} = \kappa \left( \tau_{\mu\nu} + \partial_\alpha (s_{\mu\nu}{}^{\alpha} + s_{\nu\mu}{}^{\alpha} - s^\alpha_{\phantom{\alpha}\mu\nu})/2\right),
\end{align}
making the function of $\tau_{\mu\nu}$ and $s^\alpha_{\phantom{\alpha}\mu\nu}$ absolutely clear: these tensors do not simply constitute a passive ``kinematical'' description of gravitational energy-momentum and spin, they actively determine the field's \emph{dynamics}. 

Despite the satisfying simplicity of this result, equation (\ref{Gh=tau+s}) is clearly not the best point at which to end our investigation. Having extolled the virtues of a formalism which keeps spin separate from energy-momentum, our real goal must be to find a formulation of gravity in which $\tau_{\mu\nu}$ and $s^\alpha_{\phantom{\alpha}\mu\nu}$ appear as source-terms in \emph{separate} gravitational field equations. It should come as no surprise that  Einstein-Cartan theory will provide precisely the environment in which to achieve this objective.

\subsection{The Einstein-Cartan Equations}\label{ECsource}
We will now disentangle the spin and energy-momentum in
 equation (\ref{Gh=tau+s}), formulating a quadratic approximation to Einstein-Cartan gravity in which $\tau_{\mu\nu}$ and $s^\alpha_{\phantom{\alpha}\mu\nu}$ appear as \emph{separate} source-terms in the field equations. In close analogy with the previous section, we proceed by expanding the field equations (\ref{ECFE}) to second order in $f_{\mu\nu}$ and $w_\alpha^{\phantom{a}\mu\nu}$, where
\begin{subequations}\label{deffw}
\begin{align}
\phi^*e^a_\mub&=\delta^a_\mu - f^a_{\phantom{a}\mu}/2,\\
\phi^* \omega_a^{\phantom{a}\mub\nub} &= w_a^{\phantom{a}\mu\nu},
\end{align}
\end{subequations}
are initially considered to be true to all orders; we then perform a non-linear field redefinition,
\begin{subequations}\label{redeffw}
\begin{align}\label{redeff}
f_{\mu\nu}&\to f_{\mu\nu} + O(f^2), \\\label{redefw}
w_\alpha^{\phantom{\alpha}\mu\nu}& \to w_\alpha^{\phantom{\alpha}\mu\nu} + O(f^2),
\end{align}
\end{subequations}
to generate the field equations we desire. In order to identify $\tau_{\mu\nu}$ and $s^\alpha_{\phantom{\alpha}\mu\nu}$ in these equations, it will also be necessary to express tensors of the form $w \partial f + w^2$ in terms of $h_{\mu\nu}$. To this end, we will evaluate these tensors on torsion-free perturbations (\ref{wsolve}),
\begin{align}\label{wsolve0}
w_\alpha^{\phantom{\alpha}\mu\nu}= (\partial^{[\nu}f^{\mu]}_{\phantom{\mu}\ \alpha} + \partial^{[\nu}f_{\alpha}^{\phantom{\alpha}\mu]} + \partial_\alpha f^{[\nu\mu]})/2 + O(f^2), 
\end{align}
in the ``symmetric'' rotation gauge (\ref{symgauge}):
\begin{align}\label{symgauge0}
f_{[\mu\nu]}=O(f^2).
\end{align}
These relations allow us to identify 
\begin{subequations}\label{ECh}
\begin{align}\label{f=h}
f_{\mu\nu} &\equiv h_{\mu\nu} + O(f^2),\\\label{w=dh}
w_\alpha^{\phantom{\alpha}\mu\nu}&\equiv \partial^{[\nu}h^{\mu]}_{\phantom{\mu}\alpha} + O(f^2),
\end{align}
\end{subequations}
and thus convert the quadratic parts of the field equations into the corresponding tensors of perturbative general relativity: $w \partial f+ w^2 = \partial h\partial h + O(f^3)$.\footnote{The validity of the formulae (\ref{newtensors}) can only be guaranteed in the generally-relativistic regime, i.e.\ torsion-free gravity described by a symmetric tensor field $h_{\mu\nu}$. In the linearized Einstein-Cartan theory, this corresponds to the restriction (\ref{wsolve0}) for $w_\alpha^{\phantom{\alpha}\mu\nu}$ and (\ref{symgauge0}) for  $f_{\mu\nu}$. If we already knew how to generalize $\tau_{\mu\nu}$ and $s^\alpha_{\phantom{\alpha}\mu\nu}$ beyond the generally-relativistic regime (where the gravitational field is described by unconstrained $f_{\mu\nu}$ and $w_\alpha^{\phantom{\alpha}\mu\nu}$) then it would be possible to recognise these tensors in the field equations without making such restrictions. The conversion $\partial h \to \partial f + w$ is not unique, however, so it is not immediately clear how this generalization should be achieved.}

To begin, let us focus our attention on the first Einstein-Cartan field equation (\ref{FEe}). Following the approach of section \ref{Esource}, we express the vacuum field equation in terms of the mixed Einstein tensor density,
\begin{align}\label{MixedECFE}
e G_a^{\phantom{a}b}=0,
\end{align}
where $G_a^{\phantom{a}b}\equiv(R_a^{\phantom{a}\mub}- e_a^{\mub} R/2)e_{\mub}^b $ is a function of the physical tetrad $e^a_{\mub}$ and physical spin connection $\omega_a^{\phantom{a}\mub\nub}$. Mapping this equation to the background, we expand to quadratic order in $f_{\mu\nu}$ and $w_{\alpha}^{\phantom{\alpha}\mu\nu}$, and simplify the resulting equation by taking the trace-reverse:
\begin{align}\nonumber
2 \partial_{[\mu}w_{\alpha]\nu}{}^{\alpha}&= - f \partial_{[\mu}w_{\alpha]\nu}{}^\alpha +f_{\nu\beta} \partial_{[\mu}w_{\alpha]}{}^{\beta\alpha}\\\label{1}&\quad {} + f^\alpha_{\phantom{\alpha}\beta} \partial_{[\mu}w_{\alpha]\nu}{}^{\beta} - 2w_{[\mu|}{}^{\alpha}_{\phantom{\alpha}\nu} w_{|\beta]}{}^{\beta}_{\phantom{\beta}\alpha}.
\end{align}
We now redefine $w_\alpha^{\phantom{\alpha}\mu\nu}$ according to the replacement
\begin{align}\nonumber
w_\alpha^{\phantom{\alpha}\mu\nu}&\to w_\alpha^{\phantom{\alpha}\mu\nu}  - f w_\alpha^{\phantom{\alpha}\mu\nu}/2 + f^{\beta [\mu}w_{\alpha\beta}{}^{\nu]}\\\label{wchange}&\quad {} +  f_\beta{}^{[\mu}\partial _\alpha f^{\nu]\beta}/ 4;
\end{align}
the quadratic field equation (\ref{1}) then becomes
\begin{align}\nonumber
2 \partial_{[\mu}w_{\alpha]\nu}{}^{\alpha}&= \partial_{[\mu}f w_{\alpha]\nu}{}^\alpha -\partial_{[\mu|}f_{\nu\beta} w_{|\alpha]}{}^{\beta\alpha} - \partial_{[\mu}f^{\alpha\beta} w_{\alpha]\nu\beta} \\\nonumber&\quad {} - 2w_{[\mu|}{}^{\alpha}_{\phantom{\alpha}\nu} w_{|\beta]}{}^{\beta}_{\phantom{\beta}\alpha}- \partial_{[\mu|}f^\beta_{\phantom{\beta}\nu} \partial_{|\alpha]}f^\alpha_{\phantom{\alpha}\beta}/4  \\&\quad {}+ \partial_{[\mu}f^{\beta\alpha} \partial_{\alpha]}f_{\nu\beta}/4.
\end{align}
Applying (\ref{ECh}) to the terms on the right-hand side and taking the trace-reverse of the equation, we obtain
\begin{align}\label{dw=tau}
2\partial_{[\mu} w_{\alpha] \nu}{}^{\alpha} - \eta_{\mu\nu} \partial_\alpha w_{\beta}^{\phantom{\beta}\alpha \beta} &= \kappa \tau_{\mu\nu}.
\end{align}
This is the field equation we had hoped to construct, mirroring the structure of the linearized non-vacuum field equation (\ref{EClin2a}) with gravitational energy-momentum $\tau_{\mu\nu}$ taking the place of the material energy-momentum tensor $T_{\mu\nu}$.

We now turn to the second vacuum Einstein-Cartan field equation (\ref{FEomega}). Writing this as
\begin{align}\label{Torsionzero}
\mc{T}^a_{\phantom{a}\mub\nub}=0
\end{align}
in the physical spacetime, we once again use the standard definitions (\ref{deffw}) and expand to second order in the background:
\begin{align}\nonumber
\partial_{[\mu}f^\alpha{}_{\nu]} + 2 w_{[\mu}{}^{\alpha}{}_{\nu]}&= f_{\beta[\mu}\partial^\beta f^\alpha{}_{\nu]}/2 + w_{\beta}^{\phantom{\beta}\alpha}{}_{[\nu}f^\beta{}_{\mu]}\\\label{2}&\quad {} + w_{[\mu}{}^\beta{}_{\nu]}f^{\alpha}_{\phantom{\alpha}\beta}.
\end{align}
Consistency with the first field equation (\ref{dw=tau}) requires us to redefine $w_{\alpha}^{\phantom{\alpha}\mu\nu}$ as before (\ref{wchange}) but places no constraint on the definition of $f_{\mu\nu}$; we are therefore free to make the replacement
\begin{align}\label{fredef}
f_{\mu\nu}\to f_{\mu\nu} -  f_{ \mu\alpha}f^{\alpha}_{\phantom{\alpha}\nu}/4.
\end{align}
Appling these redefinitions to the second field equation (\ref{2}) and converting the quadratic terms using (\ref{ECh}), one finds that
\begin{align}\nonumber
\partial_{[\mu}f^\alpha{}_{\nu]} + 2 w_{[\mu}{}^{\alpha}{}_{\nu]}&= 2\bar{h}_{\beta[\nu}\partial^{[\alpha} h_{\mu]}{}^{\beta]}\\
&= \kappa (s^\alpha_{\phantom{\alpha}\mu\nu} +\delta_{[\mu}{\!\!\!}^\alpha s^{\beta}{}_{\nu]\beta}),
\end{align}
from which the desired equation follows:
\begin{align}
\nonumber
 \partial_{[\mu}f^\alpha{}_{\nu]}+ 2 w_{[\mu}{}^\alpha{}_{\nu]}\hspace{2.35cm}
\\{} +\delta^\alpha_{[\mu|} (\partial_{|\nu]}f - \partial_\beta f^\beta{}_{|\nu]} -  2w_{\beta\phantom{\beta}|\nu]}^{\phantom{\beta}\beta}) &= \kappa s^\alpha_{\phantom{\alpha}\mu\nu}.
\end{align} 
We have found a suitable counterpart to equation (\ref{dw=tau}), in which the gravitational spin tensor $s^\alpha_{\phantom{\alpha}\mu\nu}$ takes on the role played by material spin $S^\alpha_{\phantom{\alpha}\mu\nu}$ in the linearized field equation (\ref{EClin2b}).

Combining these results, we conclude that under the perturbative expansions 
\begin{subequations}\label{newfw}
\begin{align}\label{newf}
\phi^*e^a_{\mub}&=\delta^a_{\mu} - f^a_{\phantom{a}\mu}/2 + f^{a}_{\phantom{a}\nu} f^\nu_{\phantom{\nu}\mu}/8,
\\
\nonumber
\phi^* \omega_a^{\phantom{a}\mub\nub} &= w_a^{\phantom{a}\mu\nu}  - f w_a^{\phantom{a}\mu\nu}/2 + w_{a\beta}{}^{[\nu}f^{\mu]\beta }\\&\quad {} +  f_\beta{}^{[\mu}\partial _a f^{\nu]\beta}/ 4,
\end{align}
\end{subequations}
the vacuum Einstein-Cartan field equations are approximated, to quadratic order, by
\begin{subequations}\label{ECquad}
\begin{align}\label{ECquada}
2\partial_{[\mu} w_{\alpha] \nu}{}^{\alpha} - \eta_{\mu\nu} \partial_\alpha w_{\beta}^{\phantom{\beta}\alpha \beta} &= \kappa \tau_{\mu\nu},\\\nonumber
 \partial_{[\mu}f^\alpha{}_{\nu]}+ 2 w_{[\mu}{}^\alpha{}_{\nu]}\hspace{2.35cm}
\\\label{ECquadb}{} +\delta^\alpha_{[\mu|} (\partial_{|\nu]}f - \partial_\beta f^\beta{}_{|\nu]} -  2w_{\beta\phantom{\beta}|\nu]}^{\phantom{\beta}\beta}) &= \kappa s^\alpha_{\phantom{\alpha}\mu\nu},
\end{align} 
\end{subequations}
wherein $\tau_{\mu\nu}$ and $s^\alpha_{\phantom{\alpha}\mu\nu}$ generate the gravitational fields $f_{\mu\nu}$ and $w_{\alpha}^{\phantom{\alpha}\mu\nu}$ in an identical fashion to the energy-momentum and spin of \emph{matter} (\ref{EClin2}). We have found the analogue of (\ref{Gh=t}) in the Einstein-Cartan theory of gravity, in which gravitational energy-momentum and spin act as the source terms of \emph{separate} field equations.

\section{Field Redefinition}\label{Redef}
We have succeeded in demonstrating that $\tau_{\mu\nu}$ and $s^\alpha_{\phantom{\alpha}\mu\nu}$ do indeed express the dynamical ``weight'' of the gravitational field, and have uncovered the field definitions, (\ref{gnew}) and (\ref{newfw}), which make this relationship manifest at the level of the field equations. We now turn our attention to the field definitions themselves, investigating the importance of (\ref{gnew}) and (\ref{newfw}) in a broader context, and exploring the effects of field redefinitions in general. In particular, the analysis of section \ref{FieldRedefs and Supers} will allow us to confirm the statement that justified the \emph{traceless condition}, which we imposed when deriving the formula for $s^\alpha_{\phantom{\alpha}\mu\nu}$ in our previous paper \cite{Ang1}.

\subsection{The ``Central'' Expansion}
In many respects, the most striking aspect of the new definition of $h_{\mu\nu}$, as displayed in (\ref{gnew}), is how closely it relates to a linear perturbation in the metric (\ref{gstandard}) and a linear  perturbation in the inverse metric (\ref{ginverse}); this is in comparison with the full range of  local Lorentz-covariant field definitions consistent with (\ref{gstandard}) at linear order: 
\begin{align}\nonumber
\phi^* g_{ab} &= \gb_{ab}+ h_{ab} + B_1 h_{ac}h^c_{\phantom{c}b} + B_2 h_{ab}h  \\\label{genredef}
&\quad  {} + \gb_{ab}(B_3 h^2 + B_4 h_{cd}h^{dc}) + O (h^3),
\end{align}
where $\{B_n\}$ are arbitrary constants. One can argue, in fact, that the definition (\ref{gnew}) lies at a natural ``center'' of the four-dimensional space parameterized by $\{B_n\}$. This argument begins by observing that \emph{a priori} there is no special variable which represents the ``true'' dynamical field of general relativity: one can equally well define the gravitational field $h_{\mu\nu}$ as a linear perturbation in a metric density $(-g)^\lambda g_{ab}$, or an inverse metric density $(-g)^\lambda g^{ab}$, for any value of $\lambda$. Of all these choices, perturbations in the metric and its inverse (i.e.\ $\lambda=0$) are distinguished by the fact that they possess linear-order gauge-transformation of the form $\partial_{(\mu}\xi_{\nu)}$, without a part proportional to $\eta_{\mu\nu} \partial_\alpha \xi^{\alpha}$, and can therefore be identified with the Fierz-Pauli massless spin-2 field. However, once we have restricted our interest to these particular definitions ((\ref{gstandard}) or (\ref{ginverse})) the decision to focus on one, and discard the other, is completely arbitrary. Instead of making a forced choice between two essentially equivalent options, one might instead consider the definition that lies exactly \emph{half-way} between them, where the values of $\{B_n\}$ are the mean of those in (\ref{gstandard}) and (\ref{ginverse}). It is easy to see that this ``center point''  is precisely the field (\ref{gnew}) that casts $t_{\mu\nu}$ as the self-interaction term of the quadratic field equations (\ref{Gh=t})! This is an extraordinary coincidence, as $\tau_{\mu\nu}$ and $s^\alpha_{\phantom{\alpha}\mu\nu}$ were selected for their capacity to display positive energy-density, causal energy-flux, and spatial spin; none of these criteria would be expected to determine a definition of $h_{\mu\nu}$ that is geometrically distinguished in this way.

\subsection{Expansion of the Einstein-Hilbert Lagrangian}\label{EHexp}
The new metric expansion (\ref{gnew}) also offers some perspective on the  particular form of the Lagrangian (\ref{Ldef}) that generated the main results of section \ref{Lag}, including the generalized formulae (\ref{newtensors}) for $\tau_{\mu\nu}$ and $s^\alpha_{\phantom{\alpha}\mu\nu}$. 

In an earlier paper \cite{Butcher09} we studied the expansion of the Einstein-Hilbert Lagrangian,\footnote{As it was Hilbert alone who formulated general relativity  in terms of a least-action principle \cite{EHaction}, some authors refer to (\ref{bbb}) as the Hilbert Lagrangian.}
\begin{align}\label{bbb}
\mc{L}_\mathrm{EH}= -\sqrt{-g}R/\kappa,
\end{align}
 under a linear perturbation of the inverse metric: $h^\prime_{\mu\nu}$ as defined in (\ref{ginverse0}).\footnote{The field written as $h_{\mu\nu}$ in \cite{Butcher09} is in fact $-h^\prime_{\mu\nu}$ in our present notation; the Lagrangians of that paper also take the opposite sign to those here.} Taking care to retain all terms proportional to background curvature, but ignoring surface terms, the Einstein-Hilbert Lagrangian was found to expand as follows,
\begin{align}\nonumber
\mc{L}_{\mathrm{EH}}[\gb^{ab} -h^{\prime ab}]&= -\sqrt{-\gb}\check{R}/\kappa + \mc{L}^\prime_1[h^{\prime ab}] +\mc{L}^\prime_2[h^{\prime ab}]\\\label{Bootexpand}&\quad{} + O(h^{\prime3}),
\end{align}
where the linear and quadratic parts of the Lagrangian, $\mc{L}^\prime_1$ and $\mc{L}^\prime_2$, are given by equations (65) and (66) of \cite{Butcher09}. 

To generate the expansion of $\mc{L}_{\mathrm{EH}}$ under our newly defined perturbation $h_{\mu\nu}$, we need only compare its metric expansion (\ref{gnew}) to that of $h^\prime_{\mu\nu}$ (\ref{ginverse}), 
\begin{align}\label{hprimetoh}
h_{ab}^\prime = h_{ab} - h_{ac}h^{c}_{\phantom{c}b}/2 + O(h^3),
\end{align}
and substitute this relation into (\ref{Bootexpand}). After commuting derivatives, and discarding surface terms, the quadratic part of this expansion turns out to be
\begin{align}\nonumber
\mc{L}_2[h^{ab}] &\equiv \mc{L}^\prime_2[h^{ab}] + \mc{L}^\prime_1[-h^{ac}h_{c}^{\phantom{c}b}/2]
\\\nonumber
& =  \frac{\sqrt{-\gb}}{4\kappa}\left(\delb_c h^{\prime a b} \delb^c \hb^\prime_{ab} - 2\delb^a \hb^\prime_{ab}\delb^c \hb^{\prime\phantom{c}b}_{\phantom{\prime}c}\right. \\
{}&\qquad\qquad \ {} +
\left.2 \hb^{\prime ab}\check{R}_{cabd}\hb^{\prime cd} \right),
\end{align}
which is precisely the form of the Lagrangian $\mc{L}$ that we used to reproduce the formulae for $\tau_{\mu\nu}$ and $s^\alpha_{\phantom{\alpha}\mu\nu}$! Thus, the curvature term in (\ref{Ldef}), which we introduced in section \ref{Lag} as an ansatz, can be understood as a consequence of the special definition of $h_{\mu\nu}$ associated with our framework: these are simply the terms proportional to the background curvature that appear when the Einstein-Hilbert action is expanded to quadratic order.

\subsection{Field Redefinitions and Superpotentials}\label{FieldRedefs and Supers}
To understand the new metric expansion (\ref{gnew}) in a wider context, we should also explain the relationship between field redefinitions and the superpotentials we encountered in section \ref{supers}.

Recall that the addition of superpotentials to $\tau_{\mu\nu}$ and $s^\alpha_{\phantom{\alpha}\mu\nu}$ corresponds to the addition of curvature terms (\ref{LRdef}) and torsion terms (\ref{torsionterms}) to the Lagrangian. Notice, however, that $t_{\mu\nu}$ is unaffected by torsion terms: the energy-momentum superpotentials (\ref{supertau}) cancel those of the spin tensor (\ref{supertaus}) when they enter the formula (\ref{Beldef}). As a result, the superpotentials of the Belinfante tensor are characterized by curvature terms alone, which were determined by the five parameters $\{A_n\}$ of equation (\ref{Sigma}). Although these parameters are fixed according to (\ref{An=}), for the sake of argument let us relax these equations and alter each $A_n$ by an amount $\Delta A_n$; the gravitational Belinfante tensor then gains the superpotential term
\begin{align}\nonumber
\kappa \Delta t_{\mu}^{\phantom{\mu}\nu} &= \kappa \partial_\alpha (\Delta s_\mu^{\phantom{\mu}\nu\alpha} + \Delta s^{\nu\phantom{\nu}\alpha}_{\phantom{\nu}\mu} -\Delta s^{\alpha\phantom{\mu}\nu}_{\phantom{\alpha}\mu})/2\\\label{supert}
&=\partial_\alpha\partial^\beta \Delta \Sigma^{\alpha\nu}{}_{\beta\mu}, 
\end{align}
where, according to (\ref{Sigma}),
\begin{align}\nonumber
\Delta \Sigma^{\alpha\beta}{}_{\mu\nu}&= \Delta A_1 h^\alpha_{[\mu}h_{\nu]}^\beta + \Delta A_2 h h^{[\alpha}_{[\mu}\delta^{\beta]}_{\nu]}+\Delta A_3 h^\gamma_{[\mu}\delta_{\nu]}^{[\beta}h^{\alpha]}_\gamma \\\label{DeltaSigma}
&\quad {} + \delta^{\alpha}_{[\mu}\delta^{\beta}_{\nu]}\left(\Delta A_4 h^2 + \Delta A_5 h_{\gamma\delta}h^{\gamma\delta}\right).
\end{align}

It is also possible to generate superpotentials in the quadratic approximation to Einstein's field equations: an arbitrary field redefinition
\begin{align}\label{Deltah}
h_{\mu\nu} \to h_{\mu\nu} + \Delta h_{\mu\nu},
\end{align}
adds the divergence-free tensor 
\begin{align}
- \widehat{G}_{\mu\nu}^{\phantom{\mu\nu} \alpha\beta}\Delta h_{\alpha\beta},
\end{align}
to the right-hand side of the field equations (\ref{Gh=t}) and thus defines a 
new Belinfante tensor,
\begin{align}
\kappa t^\prime_{\mu\nu}\equiv \kappa t_{\mu\nu } - \widehat{G}_{\mu\nu}^{\phantom{\mu\nu} \alpha\beta}\Delta h_{\alpha\beta},
\end{align}
that acts as the source of the new gravitational field. Therefore, as long as we can find a field redefinition $\Delta h_{\mu\nu} $ to solve
\begin{align}\label{superdh}
 - \widehat{G}_{\mu}^{\phantom{\mu}\nu \alpha\beta}\Delta h_{\alpha\beta}= \partial_\alpha\partial^\beta \Delta \Sigma^{\alpha\nu}{}_{\beta\mu}, 
\end{align}
we can produce the same superpotential in the field equations as the ones we have generated by altering $\{A_n\}$ in the Lagrangian.

First we shall try to solve equation  (\ref{superdh}) using \emph{local} field redefinitions. Noting that $\Delta h_{\mu\nu}$ will also need to be Lorentz-covariant and quadratic in $h_{\mu\nu}$ to solve this equation, the most general field redefinition we need to consider is
\begin{align}\nonumber
\Delta h_{\mu\nu} &= \Delta B_1 h_{\mu\alpha}h^\alpha_{\phantom{\alpha}\nu} + \Delta B_2 h_{\mu\nu}h  \\\label{genredef2}
&\quad  {} + \eta_{\mu\nu}(\Delta B_3 h^2 + \Delta B_4 h_{\alpha\beta}h^{\alpha\beta}),
\end{align}
where the $\{\Delta B_n\}$ correspond to changes in the parameters $\{B_n\}$ of equation (\ref{genredef}). Comparing the number of free parameters here with those of (\ref{DeltaSigma}) it is immediately clear that these local redefinitions will not span the entire space of superpotentials. Indeed, if we use the identity $\widehat{G}_{\mu}^{\phantom{\mu}\nu \alpha\beta}X_{\alpha\beta} \equiv -2 \partial_\alpha \partial^\beta (\delta^{[\nu}_{[\mu} \bar{X}^{\alpha]}_{\beta]})$, valid for any symmetric tensor $X_{\mu\nu}$, we can rewrite (\ref{superdh}) as
\begin{align}\label{partial}
0= \partial_\alpha\partial^\beta\left( \Delta \Sigma^{\alpha\nu}{}_{\beta\mu} -2\delta^{[\nu}_{[\mu} \Delta \bar{h}^{\alpha]}_{\beta]}\right);
\end{align}
inserting (\ref{genredef2}) and (\ref{DeltaSigma}), we arrive at 
\begin{align}
0&= \partial_\alpha\partial^\beta \!\left[ \Delta A_1 h^\alpha_{[\beta}h_{\mu]}^\nu + (\Delta A_2- 2\Delta B_2) h h^{[\alpha}_{[\beta}\delta^{\nu]}_{\mu]}\right.\\\nonumber
&\qquad\qquad {}+ (\Delta A_3-2\Delta B_1 ) h^\gamma_{[\beta}\delta_{\mu]}^{[\nu}h^{\alpha]}_\gamma \\\nonumber
&\qquad\qquad {} + \delta^{\alpha}_{[\beta}\delta^{\nu}_{\mu]}(\Delta A_4 +2\Delta B_3 + \Delta B_2 ) h^2 \\\nonumber
&\qquad\qquad \left. \!{} + \delta^{\alpha}_{[\beta}\delta^{\nu}_{\mu]}(\Delta A_5 + 2 \Delta B_4 + \Delta B_1) h_{\gamma\delta}h^{\gamma\delta}\right],
\end{align}
which makes the mismatch of parameters unequivocal. Clearly, this equation can only hold for all $h_{\mu\nu}$ if
\begin{align}\label{DeltaA1}
\Delta A_1=0,
\end{align}
and if we assume that this is the case, the local field redefinition $\Delta h_{\mu\nu}$ is determined uniquely:
\begin{align}
 \Delta B_1 &= \Delta A_3/2,&\Delta B_2 &= \Delta A_2/2,\\\nonumber
\Delta B_3&= -\Delta A_4/2 -\Delta A_2/4,&
\Delta B_4 &= -\Delta A_5/2 - \Delta A_3/4.
\end{align}
As the $\Delta A_n$ were defined relative to the values (\ref{An=}) that correspond to our Belinfante tensor $t_{\mu\nu}$, the condition (\ref{DeltaA1}) implies that
\begin{align}
A_1=-1,
\end{align}
which also arose in section \ref{FPLagsec} as the requirement that ensured the curvature terms would extend the spin-2 gauge invariance of the Lagrangian beyond the flat background. 

If $A_1\ne-1$ then it is still possible to solve equation (\ref{superdh}) by inverting the differential operator $\widehat{G}_{\mu}^{\phantom{\mu}\nu\alpha\beta}$:
\begin{align}\label{genericsolve}
\Delta \bar{h}_{\mu}^{\phantom{\mu}\nu} =  \frac{2}{\partial^2} \partial_\alpha\partial^\beta \Delta \Sigma^{\alpha\nu}{}_{\beta\mu},
\end{align}
where the precise form of the propagator $1/\partial^2$ will depend on boundary conditions. As we have seen, however, this $\Delta h_{\mu\nu}$ cannot be a \emph{local} function of $h_{\mu\nu}$; hence it will no longer be possible to express the physical metric $\phi^*g_{ab}$ as a local function of the gravitational field (\ref{genredef}). 

From this vantage point, we can now appreciate another important property of our framework: the Belinfante tensor $t_{\mu\nu}$ (constructed from $\tau_{\mu\nu}$ and $s^\alpha_{\phantom{\alpha}\mu\nu}$) is the source of a gravitational field $h_{\mu\nu}$ of which the metric is a \emph{local} function (\ref{gnew}). Of course, this was precisely the reason for fixing $A_1=-1$ (by imposing the \emph{traceless condition}) when we derived the formula for $s^\alpha_{\phantom{\alpha}\mu\nu}$ in our previous paper \cite{Ang1}. Note that the statement that we made in section IIIB of that paper, justifying $A_1=-1$, has now been proven. 

This state of affairs should be contrasted with that of the celebrated tensor $t^\mathrm{LL}_{\mu\nu}$ of Landau and Lifshitz \cite{LL}:
\begin{align}
\kappa t_\mathrm{LL}^{ab} &\equiv - \phi^* G^{ab} +  \frac{1}{\phi^*g}\delb_{c}\delb_d\!\left(\phi^*(gg^{a[b}g^{c]d})\right).
\end{align}
The divergence on the right-hand side clearly contributes a term
\begin{align}\label{LLsuper}
\partial_\alpha \partial_\beta (h^{\mu[\nu}h^{\alpha]\beta}),
\end{align}
at second order, corresponding to a superpotential with $\Delta A_1=1 \ \Rightarrow \ A_1 =0$. Thus, there can be no local field redefinition that will render the Landau-Lifshitz tensor as the source term of the vacuum Einstein field equations, and furthermore, the tensor cannot be derived from a covariantized Lagrangian which maintains its spin-2 gauge invariance beyond the flat background.

It is rather surprising that this deficiency is not more widely known. In their effort to construct a gravitational energy-momentum tensor that was symmetric and free of second derivatives \emph{in all gauges}, it seems that Landau and Lifshitz were forced to include a superpotential (\ref{LLsuper}) that would be impossible to generate in the field equations by a local redefinition of $h_{\mu\nu}$. The only way the Landau-Lifshitz tensor can be given an equal footing with the energy-momentum of matter, generating gravity alongside $T_{\mu\nu}$ in the Einstein field equations, would be to define the gravitational field $h_{\mu\nu}$ in terms of a \emph{non-local} perturbation in the metric.

\subsection{Beyond Second Order}\label{beyondso}
It will be difficult to gain any further insight into the physical meaning of the new $h_{\mu\nu}$ without first deciding how the definition (\ref{gnew}) should extend beyond quadratic order. As this question is intimately related to the issue of extending the formulae of $\tau_{\mu\nu}$ and $s^\alpha_{\phantom{\alpha}\mu\nu}$ to the full non-linear theory, we shall postpone a thorough investigation of this topic for another article. For now, we only mention one particularly attractive possibility. As we have already explained, the new definition (\ref{gnew}) lies on a point of symmetry between a linear perturbation in the metric (\ref{gstandard}) and a linear perturbation in the inverse metric (\ref{ginverse}); reflecting this, one finds that the expansion of the inverse metric, consistent with (\ref{gnew}) is 
\begin{align}\label{gnewinverse}
\phi^*g^{ab} =\gb^{ab} - h^{ab} + h^{ac}h_{ c}^{\phantom{ c}b}/2 + O(h^3),
\end{align}
which, apart from a change in sign convention $h_{\mu\nu}\to - h_{\mu\nu}$, is identical to the metric expansion (\ref{gnew}) to quadratic order. Thus, a particularly natural extension of (\ref{gnew}) would be one which preserved this symmetry \emph{exactly}, so that the metric and its inverse had an identical expansion to all orders, except for a sign change in $h_{\mu\nu}$. This idea can be realized by viewing the tensor $h^a_{\phantom {a}b}/2$ as a linear map and forming its exponential $[e^{h/2}]^a_{\phantom{a}b}$; a metric defined by
\begin{align}\label{gexp}
\phi^* g_{ab} \equiv [e^{h/2}]^c_{\phantom{c}a} \gb_{cd} [e^{h/2}]^d_{\phantom{d}b},
\end{align}
is then consistent with (\ref{gnew}) to second order, and moreover, the associated expansion of the inverse metric,
\begin{align}
\phi^* g^{ab} = [e^{-h/2}]^a_{\phantom{a}c} \gb^{cd} [e^{-h/2}]^b_{\phantom{c}d} ,
\end{align}
is clearly identical to (\ref{gexp}) apart from a change in the sign of $h_{\mu\nu}$. 

\subsection{New Fields for Einstein-Cartan}\label{ECredefs}
While it is certainly tempting to bring our analysis to bear on the new field definitions (\ref{newfw}) that arose in Einstein-Cartan theory, unfortunately a full discussion of these variables will not be possible at this time. As we previously explained, the right-hand sides of (\ref{ECquad}) are only given by the formulae (\ref{newtensors}) when we restrict ourselves to the torsion-free perturbations (\ref{wsolve0}) and symmetric gravitational field (\ref{symgauge0}) of general relativity. We are therefore free to alter the quadratic parts of the new field definitions (\ref{newfw}) by terms proportional to $\mc{T}_{\alpha}{}^{\mu\nu}$ and $f_{[\mu\nu]}$: such terms vanish under the aforementioned restrictions, and so do not interfere with our results. Until we fix the definitions of $\tau_{\mu\nu}$ and $s^\alpha_{\phantom{\alpha}\mu\nu}$ in terms of arbitrary perturbations $f_{\mu\nu}$ and $w^\alpha_{\phantom{\alpha}\mu\nu}$, this degeneracy will remain, and it will be difficult to offer a physical interpretation of the new field variables. We leave this generalization of $\tau_{\mu\nu}$ and $s^\alpha_{\phantom{\alpha}\mu\nu}$, and the consequent analysis of the field variables they define, for another paper. For now, we shall simply remark that the new tetrad expansion (\ref{newf}) defines an inverse tetrad
\begin{align}
\phi^*e_a^\mub=\delta_a^\mu + f^\mu_{\phantom{\mu}a}/2 + f^\mu_{\phantom{\mu}\nu}f^\nu_{\phantom{\nu}a} /8 + O(f^3),
\end{align}
which, to second order, differs from the tetrad expansion only by a change in sign convention $f_{\mu\nu}\to -f_{\mu\nu}$: mirroring the relationship between the metric expansion (\ref{gnew}) and inverse metric expansion (\ref{gnewinverse}) for the new definition of $h_{\mu\nu}$ that arose in section \ref{Esource}. In fact, $f_{\mu\nu}$ defines a metric 
\begin{align}\nonumber
\phi^* g_{ab} &= \gb_{ab} + f_{(ab)} + f_{(a|}{}^\mu f_{\mu |b)}/4 +  f^\mu_{\phantom{\mu} a} f_{\mu b}/4\\&\quad {}  + O(f^3),
\end{align}
which, if we set
\begin{align}\label{f=hquad}
f_{\mu\nu}= h_{\mu\nu} + O(f^3),
\end{align}
gives
\begin{align}
\phi^* g_{ab} &= \gb_{ab} + h_{ab} + h_{a}{}^\mu h_{\mu b}/2  + O(h^3),
\end{align}
identical to the new metric expansion (\ref{gnew}) of section \ref{Esource}. Note that the equivalence (\ref{f=hquad}) between $f_{\mu\nu}$ and $h_{\mu\nu}$ now holds to \emph{quadratic} order, whereas one would only expect a linear correspondence from the symmetric gauge (\ref{f=hlin}) condition alone.

\section{Conclusion}
As a local description of the energy-momentum and spin of the linearized gravitational field, $\tau_{\mu\nu}$ and $s^\alpha_{\phantom{\alpha}\mu\nu}$ serve a number of purposes within the theory. In addition to accounting for the energy-momentum and angular momentum exchanged locally with matter \cite{Butcher10, Ang1}, we can now confirm their status as Noether currents of translational and rotational symmetry, and as sources of gravity itself. Thus, our framework displays many of the fundamental properties possessed by previous treatments of local gravitational energetics \cite{Einstein, DiracPsuedo,LL}, all the while endowing linear gravity with positive energy-density, causal energy-flux, and spatial spin.

In summary, the main results of this article are as follows:
\begin{itemize}
\item A Lagrangian (\ref{Ldef}) has been constructed from which our gravitational energy-momentum tensor $\tau_{\mu\nu}$ and spin tensor $s^\alpha_{\phantom{\alpha}\mu\nu}$ arise according to standard variational definitions (\ref{tau+sdef}) once harmonic gauge (\ref{harmonic}) has been enforced.
\item The formulae for $\tau_{\mu\nu}$ and $s^\alpha_{\phantom{\alpha}\mu\nu}$ have been generalised beyond harmonic gauge (\ref{newtensors}). This is the unique generalisation that does not add terms of the form $h\partial^2 h$ to $\tau_{\mu\nu}$.
\item The Belinfante tensor associated with our description $t_{\mu\nu}\equiv \tau_{\mu\nu} + \partial_\alpha (s_{\mu\nu}^{\phantom{\mu\nu}\alpha} + s_{\nu\mu}^{\phantom{\mu\nu}\alpha} - s^\alpha_{\phantom{\alpha}\mu\nu})/2$ has been calculated (\ref{t}).
\item A non-linear perturbation in the metric has been uncovered (\ref{gnew}) which generates $\tau_{\mu\nu}$ and $s^\alpha_{\phantom{\alpha}\mu\nu}$ (combined into $t_{\mu\nu}$) as the second-order contribution to the Einstein field equations (\ref{Gh=t}).
\item Non-linear perturbations in the tetrad and spin connection have been found (\ref{newfw}) which generate $\tau_{\mu\nu}$ and $s^\alpha_{\phantom{\alpha}\mu\nu}$ \emph{separately} as the second order contributions to the Einstein-Cartan field equations (\ref{ECquad}).
\item The global quantities defined by $\tau_{\mu\nu}$ and $s^\alpha_{\phantom{\alpha}\mu\nu}$ have been examined (appendix \ref{ADM}) and shown to agree with ADM energy-momentum and angular momentum under appropriate conditions.
\end{itemize}
In addition, we have described at quadratic order:
\begin{itemize}
\item The correspondence between background coupling in the Lagrangian and the superpotentials present in the gravitational energy-momentum tensor and spin tensor that the Lagrangian defines. (Section \ref{supers})
\item The correspondence between these superpotentials and the field redefinitions required to generate those particular gravitational energy-momentum tensors and spin tensors in the field equations. (Section \ref{FieldRedefs and Supers})
\end{itemize}
Finally, as a consequence of the above analysis: 
\begin{itemize}
\item  We have isolated the range of superpotentials ($A_1\ne-1$ in (\ref{Sigma})) which (i) cannot be generated by a Lagrangian that maintains the gauge freedom of $h_{ab}$ beyond the flat background (\ref{GIsucc}) and (ii) cannot be generated in the field equations by \emph{local} field redefinitions (\ref{genredef}). The Landau-Lifshitz tensor contains a superpotential of this form; as such, non-local field redefinitions are required to render the Landau-Lifshitz tensor as a source for the gravitational field.
\end{itemize}

Having embedded $\tau_{\mu\nu}$ and $s^\alpha_{\phantom{\alpha}\mu\nu}$ within various aspects of the linear and quadratic approximations to gravity, the key goal that remains is to extend these ideas to the full non-linear theory. This is obviously an ambitious task, and at the present stage it is far from clear which of our framework's properties can survive in the exact theory. However, based on the results of this article, it seems more than likely that Einstein-Cartan gravity will provide a natural starting point from which to launch this undertaking, with the field definitions (\ref{gnew}), (\ref{gexp}) and (\ref{newfw}) offering clues as to the new field variables into which this theory should be cast. 

\begin{acknowledgments}
L.\,M.\,B.\ is supported by STFC, St.\ John's College, Cambridge, and Jesus College, Cambridge. We thank Richard Arnowitt and Stanley Deser for bringing equation (\ref{thanks}) to our attention.
\end{acknowledgments}

\appendix

\section{Einstein-Cartan Theory}\label{ECintro}
The role of this appendix is to briefly introduce Einstein-Cartan gravity, establish notation, and serve as a reference for results needed in the body of the paper. For a more complete treatment of the subject, see \cite{HehlEC, BlagEC, GTG, GTG2}.

\subsection{Kinematics}

Einstein-Cartan theory is a slight extension of general relativity, in which (as formulated by Kibble \cite{Kibble} and Sciama \cite{Sciama1,Sciama2}) translational and rotational symmetries are gauged separately, rather than being subsumed into a single diffeomorphism gauge transformation. The gravitational field is represented by four vector fields $e^a_\mub$ (the tetrad) and six covector fields $\omega_{a}^{\phantom{a}\mub\nub}=-\omega_{a}^{\phantom{a}\nub\mub}$ (the spin connection). Following the conventions of our previous papers \cite{Butcher10, Ang1} and Wald \cite{Wald}, Greek letters are used as numerical indices (running from 0 to 3, raised and lowered by $\eta_{\mu\nu}=\text{diag}(-1,1,1,1)$) while Roman indices represent the tensor ``slots'' of Penrose's abstract index notation \citep[chap.\ 2.4]{Wald}. Note that the numerical indices now come in two varieties: the unadorned Greek letters are used to enumerate the components of tensors in a Lorentzian coordinate system $\{x^\mu\}$ of flat spacetime, whereas the Greek letters with overbars enumerate the components of tensors with respect to the non-holonomic basis $\{ e^a_\mub \}$ formed from the tetrad. For example, 
\begin{align}
v_\mub & \equiv e^a_\mub v_a, &v^\mub &\equiv  e^\mub_a v^a,
\end{align}
where $e^\mub_a$ is the inverse tetrad, defined by $e^a_\mub e^\nub_a= \delta^\mub_\nub$. In order that this convention be consistent with general relativity (in which abstract indices are raised and lowered by the metric $g_{ab}$)  we identify the metric as follows:
\begin{align}\label{metricdef}
g_{ab}\equiv e^\mub_a e_{\mub b}.
\end{align}

The tetrad and the spin connection are gauge fields: they allow the global translational and rotational symmetries of flat space to be generalized to \emph{local} symmetries. Local translations are brought about by diffeomorphisms $\varphi: \mc{M}\to \mc{M}$, under which a scalar field $\psi$ has a particularly simple transformation law: $\psi(x)\to \psi(\varphi^{-1}(x))$. The tetrad allows us to form a ``translation covariant'' derivative $\partial_\mub \equiv e_\mub^a \partial_a$; as this derivative carries no spacetime indices, $\partial_\mub\psi$ will also transform as a scalar field. Local rotations are embodied by position-dependent Lorentz transformations which act on the tetrad,
\begin{align}\label{localrot}
e^a_\mub &\to e^a_\nub \Lambda^\nub_{\phantom{\nub}\mub}(x),&  \Lambda^\alphab_{\phantom{\alphab}\mub} \Lambda^\betab_{\phantom{\betab}\nub} \eta_{\alphab\betab} &=\eta_{\mub\nub}.
\end{align}
Neither $\partial_a$ nor $\partial_\mub$ transform covariantly under local rotations ($\partial_a v_\mub \to \partial_a( v_\nub \Lambda^\nub_{\phantom{\nub}\mub})=  \Lambda^\nub_{\phantom{\nub}\mub}\partial_a v_\nub + v_\nub\partial_a \Lambda^\nub_{\phantom{\nub}\mub}$) so a ``rotation covariant'' derivative $D_a$ is constructed using the spin connection: 
\begin{align}\nonumber
D_a v_\mub &= \partial_a v_\mub - \omega_{a\phantom{\alphab}\mub}^{\phantom{a}\alphab} v_\alphab,\\
D_a v^\mub &= \partial_a v^\mub + \omega_{a\phantom{\mub}\alphab}^{\phantom{a}\mub} v^\alphab,  \quad \text{etc.}
\end{align}
If we declare that the spin connection should transform according to
\begin{align}
\omega_{a\phantom{\mub}\nub}^{\phantom{a}\mub} \to (\Lambda^{-1})^\mub_{\phantom{\mub}\alphab} \left( \partial_a \Lambda^\alphab_{\phantom{\alphab}\nub} + \omega _{a\phantom{\alphab}\betab}^{\phantom{a}\alphab}\Lambda^\betab_{\phantom{\betab}\nub}  \right),
\end{align}
under local rotations, then it is easy to show that these derivatives are indeed covariant: 
\begin{align}\nonumber
D_a v_\mub &\to \Lambda^\nub_{\phantom{\nub}\mub}D_a v_\nub,\\
D_a v^\mub &\to (\Lambda^{-1})^\mub_{\phantom{\mub}\nub}D_a v^\nub,\quad \text{etc.}
\end{align}

Using the tetrad once again, we can now construct a derivative $D_\mub \equiv e^a_\mub D_a$ that is covariant under both local translations and rotations. Thus, by replacing all partial derivatives with covariant derivatives, and all spacetime indices with basis indices, we can ``gauge'' the global Poincar\'e-invariance of any flat-space field theory, and in doing so, extend the theory to a spacetime with curvature and torsion.

\subsection{Curvature and Torsion}
The rotation covariant derivatives define a curvature tensor $R_{ab \phantom{\nub} \mub}^{\phantom{ab}\nub}$: 
\begin{align}
[D_a, D_b]v_\nub \equiv- R_{ab\phantom{\mub} \nub}^{\phantom{ab}\mub} v_\mub \quad \forall v_\mub,
\end{align}
from which it follows that
\begin{align}\label{R}
R_{ab\phantom{\mub} \nub}^{\phantom{ab}\mub}= 2 \left( \partial_{[a} \omega_{b]}{}^{\mub}_{\phantom{\mub}\nub} +  \omega_{[a|}{}^{\mub}_{\phantom{\mub}\alphab} \omega_{|b]}{}^{\alphab}_{\phantom{\alphab}\nub} \right).
\end{align}
Contracting this tensor with the tetrad yields an asymmetric Ricci tensor $R_{b\nub} \equiv e^a_\mub R_{ab\phantom{\mub} \nub}^{\phantom{ab}\mub}$, and Ricci scalar $R \equiv e_\mub^a R_a^{\phantom{a} \mub}$.

The major difference between Einstein-Cartan gravity and general relativity is that, in addition to curvature, spacetime may also possess torsion, which we quantify with the torsion tensor $\mc{T}^\alphab_{\phantom{\alphab}ab} \equiv2 D_{[a} e^\alphab_{b]},$ or equivalently,
\begin{align}\label{torsiondef2}
\mc{T}^a_{\phantom{a}\mub\nub} \equiv-2 D_{[\mub} e^a_{\nub]}.
\end{align}
If this tensor vanishes everywhere, we can use $D_{[\mub}e^a_{\nub]} =0$ to relate the spin connection to the tetrad:
\begin{align}\label{omegasolve}
\omega_a^{\phantom{a}\mub\nub} = e^\mub_b \nabla_a e^{\nub b},
\end{align}
where $\nabla_a$ is the familiar (torsion-free) metric-compatible derivative of general relativity. Thus, when there is no torsion, the covariant derivatives $D_\mub$ are equivalent to $\nabla_a$ in the sense that $D_\mub v_\nub = e^a_\mub e^b_\nub \nabla_a v_b$. Thus, in the absence of torsion, the curvature tensor $R_{ab \phantom{\nub} \mub}^{\phantom{ab}\nub}$ is equivalent to its counterpart from general relativity.

\subsection{Dynamics}
The dynamics of the tetrad, spin connection, and matter fields are determined by a Lagrangian $\mc{L}_\mathrm{EC}$ that closely resembles that of the Einstein-Hilbert action: 
\begin{align}
\mc{L}_\mathrm{EC}= - e R/\kappa  + \mc{L}_\mathrm{matter},
\end{align}
where $\mc{L}_\mathrm{matter}$ is the (covariantized) matter Lagrangian and $e\equiv \det(e^\mub_a)= 1/\det(e_\mub^a)= \sqrt{-g}$ is the volume element. Variation with respect to $e_\mub^a$ and $\omega_a^{\phantom{a}\mub\nub}$ generates the gravitational field equations,
\begin{subequations}\label{ECFE}
\begin{align}\label{FEe}
G_a^{\phantom{a}\mub} & = \kappa T_{a}^{\phantom{a} \mub},\\\label{FEomega}
F^a_{\phantom{a}\mub\nub} & = \kappa S^a_{\phantom{a}\mub\nub},
\end{align}
\end{subequations}
where matter's energy-momentum tensor $T_a^{\phantom{a}\mub}$, and spin tensor $S^a_{\phantom{a}\mub\nub}$, are the conjugate currents of the translational and rotational gauge fields,
\begin{align}\label{T+Sdef}
T_a^{\phantom{a}\mub}&\equiv\frac{1}{2 e}\frac{\delta \mc{L}_\mathrm{matter}}{\delta e_\mub^a},& S^a_{\phantom{a}\mub\nub}&\equiv 
\frac{1}{e}\frac{\delta \mc{L}_\mathrm{matter}}{\delta \omega_a^{\phantom{a}\mub\nub}},
\end{align}
and we have written
\begin{subequations}
\begin{align}
G_a^{\phantom{a}\mub} &\equiv R _{a}^{\phantom{a} \mub}- e^\mub_a R/2,\\ F^a_{\phantom{a}\mub\nub}&\equiv2 e^{-1}D_b(e e_{[\mub}^a e_{\nub]}^b)= \mc{T}^a_{\phantom{a}\mub\nub} + 2 e^a_{[\mub}\mc{T}^\alphab_{\phantom{\alphab}\nub] \alphab}.
\end{align}
\end{subequations}
Consequently, the energy-momentum of matter generates curvature, and the intrinsic spin of matter generates torsion. When $S^a_{\phantom{a}\mub\nub}=0$ everywhere, the second field equation (\ref{FEomega}) ensures that torsion will vanish also; on substitution of (\ref{omegasolve}) and (\ref{metricdef}), the first field equation (\ref{FEe}) then becomes the usual Einstein field equations. 

\subsection{Perturbations}\label{ECpert}
When the curvature and torsion of the physical spacetime are small, it is often convenient to represent the gravitational fields as perturbations from a flat torsion-free background $(\bc{M},\check{e}^a_\mub,\check{\omega}_a^{\phantom{a}\mub\nub})$. This spacetime is equipped with a constant tetrad 
\begin{subequations}\label{flate}
\begin{align}\nonumber
\check{e}^a_\mub &= \delta^a_\mu\equiv (\partial/ \partial x^\mu)^a,\\
\check{e}^\mub_a &= \delta^\mu_a \equiv (\ud x^\mu)_a,
\end{align}
defined by a Lorentzian coordinate system $\{x^\mu\}$, and a vanishing spin connection,
\begin{align}\label{flatomega}
\check{\omega}_a^{\phantom{a}\mub\nub}&=0.
\end{align}
\end{subequations}
In the background spacetime it is customary to manipulate indices using the background tetrad; because $\check{e}^\mu_\nub=\delta^\mu_\nu$, the distinction between barred indices and unbarred indices can then be dropped.

Mapping the physical spacetime $(\mc{M},e^a_\mub,\omega_a^{\phantom{a}\mub\nub})$ onto the background with a diffeomorphism $\phi: \mc{M}\to\bc{M}$, we define an asymmetric tensor field $f^a_{\phantom{a}\mu}$ as a perturbation in the physical tetrad,
\begin{align}\label{deff}
\phi^*e^a_\mub=\delta^a_\mu - f^a_{\phantom{a}\mu}/2,
\end{align}
and the tensor field $w_a^{\phantom{a}\mu\nu} =-w_a^{\phantom{a}\nu\mu}$ as a perturbation in the physical spin connection,
\begin{align}\label{defw}
\phi^* \omega_a^{\phantom{a}\mub\nub} = w_a^{\phantom{a}\mu\nu}.
\end{align}
The perturbation in the tetrad (\ref{deff}) will be accompanied by a perturbation in the inverse tetrad,
\begin{align}
\phi^*e_a^\mub=\delta_a^\mu + f^\mu_{\phantom{\mu}a}/2 + f^\mu_{\phantom{\mu}\nu}f^\nu_{\phantom{\nu}a} /4 + O(f^3),
\end{align}
which in turn perturbs the physical metric (\ref{metricdef}): 
\begin{align}\nonumber
\phi^* g_{ab} &\equiv \gb_{ab} + f_{(ab)} + f_{(a|}{}^\mu f_{\mu |b)}/2 + f^\mu_{\phantom{\mu} a} f_{\mu b}/4 \\\label{metricf}
 &\quad {} + O(f^3).
\end{align}
Thus, in the linear approximation, we can identify the symmetric part of $f_{\mu\nu}$ with the metric perturbation $h_{\mu\nu}$ of general relativity:
\begin{align}
h_{\mu\nu}\equiv f_{(\mu\nu)} + O(f^2).
\end{align}

Working to first order in $f_{\mu\nu}$ and $w_a^{\phantom{a}\mu\nu}$, the Einstein-Cartan field equations (\ref{ECFE}) take the following form:
\begin{subequations}\label{EClin2}
\begin{align}\label{EClin2a}
2\partial_{[\mu} w_{\alpha] \nu}{}^{\alpha} - \eta_{\mu\nu} \partial_\alpha w_{\beta}^{\phantom{\beta}\alpha \beta} &= \kappa T_{\mu\nu},\\\nonumber
 \partial_{[\mu}f^\alpha{}_{\nu]}+ 2 w_{[\mu}{}^\alpha{}_{\nu]}\hspace{2.35cm}
\\\label{EClin2b}{} +\delta^\alpha_{[\mu|} (\partial_{|\nu]}f - \partial_\beta f^\beta{}_{|\nu]} -  2w_{\beta\phantom{\beta}|\nu]}^{\phantom{\beta}\beta}) &= \kappa S^\alpha_{\phantom{\alpha}\mu\nu},
\end{align} 
\end{subequations}
where, for the sake of notational brevity, we have dropped the $\phi^*$ from the tensors $\phi^* T_a^{\phantom{a}\mub}$ and $\phi^*S^a_{\phantom{a}\mub\nub}$. To recover the linearized field equations of general relativity, we need only set $S^\alpha_{\phantom{\alpha}\mu\nu}=0$: the solution to equation (\ref{EClin2b}) is then
\begin{align}\label{wsolve}
w_\alpha^{\phantom{\alpha}\mu\nu}= (\partial^{[\nu}f^{\mu]}_{\phantom{\mu}\ \alpha} + \partial^{[\nu}f_{\alpha}^{\phantom{\alpha}\mu]} + \partial_\alpha f^{[\nu\mu]})/2 , 
\end{align}
which can be substituted into (\ref{EClin2a}) to retrieve
\begin{align}\label{Gf=T}
\widehat{G}_{\mu\nu}^{\phantom{\mu\nu}\alpha\beta}f_{(\alpha\beta)} = \kappa T_{\mu\nu}. 
\end{align}

The equivalence between $f_{\mu\nu}$ and $h_{\mu\nu}$ can be strengthened further by fixing the rotation gauge freedom (to linear order) with the condition
\begin{align}\label{symgauge}
f_{[\mu\nu]}=O(f^2).
\end{align}
This can always be achieved by a local rotation (\ref{localrot}) of the form
\begin{align}
\Lambda^\mu_{\phantom{\nu}\nu}= \delta^\mu_\nu +(f^\mu_{\phantom{\mu}\nu}-f^{\phantom{\nu}\mu}_{\nu})/4 +O(f^2),
\end{align}
which is indeed a valid Lorentz transformation, $\Lambda^\alpha_{\phantom{\alpha}\mu} \Lambda^\beta_{\phantom{\beta}\nu} \eta_{\alpha\beta} =\eta_{\mu\nu} + O (f^2)$, and has the desired effect: $f_{\mu\nu}\to f_{(\mu\nu)} + O(f^2)$. In this ``symmetric'' gauge, $f_{\mu\nu}$ and $h_{\mu\nu}$ are equivalent to linear order:
\begin{align}\label{f=hlin}
h_{\mu\nu}= f_{\mu\nu} + O(f^2).
\end{align}

\section{An Identity}\label{Id}
Here we derive an identity that relates the Belinfante tensor (\ref{t}) to the Einstein tensor of physical spacetime. First, for notational purposes, let us define a tensor $\widetilde{G}^{(2)}{\!}_a^{\phantom{a}b}$ to represent the quadratic part of the ``mixed'' Einstein tensor density:
\begin{align}
\widetilde{G}^{(2)}{\!}_a^{\phantom{a}b}&\equiv\left[\phi^* (\sqrt{-g} G_a^{\phantom{a}b})\right]^{(2)}.
\end{align}
This definition expands to give
\begin{align}
\widetilde{G}^{(2)}_{\mu\nu}&= R^{(2)}_{\mu\nu} -  R^{(1)}_{\mu\alpha} \hb^\alpha_{\phantom{\alpha}\nu} - \eta_{\mu\nu}(R^{(2)} - R^{(1)}_{\alpha\beta}\hb^{\alpha\beta})/2,
\end{align}
where
\begin{align}
R^{(1)}_{ab} &\equiv \left[\phi^*R_{ab}\right]^{(1)},&R^{(2)}_{ab} &\equiv \left[\phi^*R_{ab}\right]^{(2)},
\end{align}
are the linear and quadratic parts of the Ricci tensor, when expanded according to (\ref{gstandard}):
\begin{subequations}\label{R2def}
\begin{align}
R^{(1)}_{\mu\nu} &=  \partial_\alpha \partial_{(\mu} h_{\nu)}{}^{\alpha} -  \partial^2 h_{\mu\nu}/2- \partial_\mu\partial_\nu h/2,
\\\nonumber
R^{(2)}_{\mu\nu}& = -  h^{\alpha\beta} (2 \partial_\alpha \partial_{(\mu}h_{\nu) \beta} - \partial_{\mu}\partial_\nu h_{\alpha\beta} - \partial_\alpha \partial_\beta h_{\mu\nu})/2 
\\\nonumber
&\quad {}+ \partial_\mu h^{\alpha\beta}\partial_\nu h_{\alpha\beta}/4 + \partial^\alpha h^\beta_{\phantom{\beta}\mu}(\partial_{[\alpha} h_{\beta]\nu} )\\ 
&\quad {}- \partial_{\alpha} \hb^{\alpha\beta}(\partial_{(\mu}h_{\nu)\beta} - \partial_\beta h_{\mu\nu}/2 ).
\end{align}
\end{subequations}
Now consider the tensor  
\begin{align}\label{Qdef}
 Q_{\mu\nu}\equiv  \widetilde{G}^{(2)}_{\mu\nu} + \frac{1}{2}  \widehat{G}_{\mu\nu}^{\phantom{\mu\nu}\alpha\beta}(h_{\alpha\gamma}h^\gamma_{\phantom{\gamma}\beta}),
\end{align}
and take its trace-reverse:
\begin{align}\nonumber
\bar{Q}_{\mu\nu}& =  R^{(2)}_{\mu\nu} - R^{(1)}_{\mu\alpha} \hb^\alpha_{\phantom{\alpha}\nu} + \frac{1}{2}\partial_\alpha \partial_{(\mu} (h_{\nu)\beta}h^{\beta \alpha} )\\ \label{Delta}&\quad {}- \frac{1}{4} \partial^2(h_{\mu\alpha}h^\alpha_{\phantom{\alpha}\nu}) - \frac{1}{4} \partial_\mu\partial_\nu (h_{\alpha\beta}h^{\alpha\beta}).
\end{align}
Substituting equations (\ref{R2def}) into (\ref{Delta}), one finds
\begin{align}\nonumber 
\bar{Q}_{\mu\nu}&= - \frac{1}{2} h^{\alpha\beta}(\partial_\alpha \partial_{(\mu} h_{\nu)\beta} - \partial_\alpha\partial_\beta h_{\mu\nu}) 
\\\nonumber
 &\quad {}- \frac{1}{2}h_{\alpha(\mu}(\partial^\alpha \partial^\beta h_{\nu) \beta} -\partial_{\nu)}\partial_\alpha h) 
\\\nonumber
 &\quad {}+ \frac{1}{4} h (2 \partial^\alpha \partial_{(\mu} h_{\nu)\alpha} - \partial^2 h_{\mu\nu} - \partial_\mu\partial_\nu h) 
\\\nonumber
 &\quad {}-\frac{1}{4}\partial_\mu h^{\alpha\beta} \partial_\nu h_{\alpha\beta} - \frac{1}{2} \partial^\alpha h^\beta_{\phantom{\beta}\mu}\partial_\beta h_{\nu \alpha}
\\\nonumber
 &\quad {} - \frac{1}{2}\partial_\alpha h^{\alpha\beta}(\partial_{(\mu}h_{\nu)\beta} -\partial_\beta h_{\mu\nu}) + \frac{1}{2} \partial_\alpha h_{\beta(\mu}\partial_{\nu)} h^{\alpha\beta}
\\\nonumber
 &\quad {}+ \frac{1}{4} \partial^\alpha h (2\partial_{(\mu}h_{\nu)\alpha} - \partial_\alpha h_{\mu\nu}) - h^{\alpha}{}_{[\nu}R^{(1)}_{\mu]\alpha}
\\\label{Deltat}
&=-\kappa \bar{t}_{\mu\nu},
\end{align}
the last line of which can be confirmed by expanding out all the trace-reversed fields on the right-hand side of (\ref{t}) and observing that $h^{\alpha}{}_{[\nu}R^{(1)}_{\mu]\alpha} = h^{\alpha}{}_{[\nu}G^{(1)}_{\mu]\alpha}=h^{\alpha}{}_{[\nu}\widehat{G}_{\mu]}{}^{\alpha\beta\gamma}h_{\beta\gamma}$. Inserting (\ref{Qdef}) into the trace-reverse of (\ref{Deltat}), we conclude that the following identity
\begin{align}
\kappa t_{\mu\nu} &= -\widetilde{G}^{(2)}_{\mu\nu} - \widehat{G}_{\mu\nu}^{\phantom{\mu\nu}\alpha\beta}(h_{\alpha\gamma}h^\gamma_{\phantom{\gamma}\beta})/2,
\end{align}
is valid for all $h_{\mu\nu}$.

\section{ADM Energy-Momentum}\label{ADM}
In this article, and those that have preceded it \cite{Butcher10, Ang1}, we have focussed on the \emph{local} aspects of gravitational energy-momentum and spin; although we will not attempt a thorough investigation here, it will be valuable to briefly examine the \emph{global} energy and momentum that our framework defines, and compare these quantities to the well-known results of Arnowitt, Deser, and Misner (ADM) \cite{ADMIII,ADMIV,ADMSummary}.

Recall that, as seen in (\ref{global}), the Belinfante tensor $t_{\mu\nu}$ defines the same \emph{total} energy, momentum, and angular momentum as $\tau_{\mu\nu}$ and $s^\alpha_{\phantom{\alpha}\mu\nu}$.\footnote{To be precise: the integrals in (\ref{global}) may in fact differ by surface terms quadratic in $h_{\mu\nu}$. However, as we will see, these can be neglected in comparison to the surface terms linear in $h_{\mu\nu}$ when the boundary of the integral is taken to spatial infinity.} Thus, for the purposes of this appendix, we are free to use whichever set of tensors is convenient, and the results we derive will carry over to the other. With this in mind, let us define the total energy-momentum of gravity and matter by
\begin{align}\label{Pdef}
\mathfrak{P}_{\mu}\equiv \int  \ud^3 y\sqrt{-g}\left(T^{\mathrm{Bel}}{}_a^{\phantom{a}b}+ (\phi^{-1})^* t_a^{\phantom{a}b}\right) (\ud/\ud y^\mu)^a (\ud y^0)_b ,
\end{align}
where $y^\mu \equiv (\phi^{-1})^*x^\mu$ are the images of the Lorentzian coordinates $\{x^\mu\}$ in the physical spacetime.\footnote{In equation (\ref{Pdef}), the contraction between the energy-momentum tensors and the covector $(\ud y^0)_a$ defines, in the usual way, the energy-momentum \emph{densities} on the surface of integration  $y^0=\text{const}$. The vectors $ (\ud/\ud y^\mu)^{a}$ have assumed the role of killing vectors in the absence of an exact spacetime symmetry; these same vectors were denoted by $e_\mu^{a}$ in \cite{Butcher10, Ang1}, but this symbol is now being used for the Einstein-Cartan tetrad.}  To this definition we now apply the Einstein field equations:
\begin{align}\nonumber
\mathfrak{P}_{\mu}&=\frac{1}{\kappa} \int  \ud^3 y\sqrt{-g}\left(G_a^{\phantom{a}b}+ \kappa (\phi^{-1})^* t_a^{\phantom{a}b}\right) (\ud/\ud y^\mu)^a (\ud y^0)_b \\
&=\frac{1}{\kappa} \int  \ud^3 x\left(\widehat{G}_\mu^{\phantom{\mu}0\alpha\beta}h_{\alpha\beta}+ \widetilde{G}^{(2)}{}_\mu^{\phantom{\mu}0}  +\kappa t_\mu^{\phantom{\mu}0} \right),
\end{align}
where, in the second line, we have evaluated the integral in terms of background quantities, expanded the metric according to (\ref{gstandard}), and neglected terms $O(h^3)$ under the assumption that the gravitational field is everywhere weak enough that the quadratic approximation to general relativity will suffice. We now use the identity (\ref{Gtid}) to write the energy-momentum as 
\begin{align}\label{P1}
\mathfrak{P}_{\mu} =\frac{1}{\kappa} \int  \ud^3 x\widehat{G}_{\mu}^{\phantom{\mu}0\alpha\beta}\left(h_{\alpha\beta} + O(h^2)\right).
\end{align}
Although terms $O(h^2)$ cannot be neglected in general (otherwise $t_{\mu\nu}$ should never have appeared in the integral (\ref{Pdef}) to begin with) we note that all the terms in (\ref{P1}) are total spatial derivatives, so $\mathfrak{P}_{\mu}$ will depend only on the behavior of the gravitational field on the boundary of the integral. Thus, as the limit is taken in which this boundary moves to spatial infinity, we will require the linear surface terms $\partial h\sim 1/r^2$ in order that integral be finite, and as a consequence the quadratic surface terms $h\partial h \sim 1/r^3$ will be negligible in comparison.

Let us first consider the total energy of the system:
\begin{align}\nonumber
\mathfrak{P}^0&=  \frac{1}{2\kappa} \int  \ud^3 x(\partial_i \partial_j h_{ij} - \partial_i\partial_i h_{jj})\\
&= \frac{1}{2\kappa} \int  \ud^2 S_i(\partial_j h_{ij} - \partial_i h_{jj}),
\end{align}
which the reader will recognise as the ADM mass \cite{ADMIV,ADMSummary}. Furthermore, the total linear momentum 
\begin{align}\nonumber
\mathfrak{P}_i&= \frac{-1}{2\kappa} \int  \ud^3 x(\partial_j\dot{h}_{ij} - \partial_k\partial_k h_{0i}- \partial_i \dot{h}_{jj} +\partial_i \partial_j h_{0j})\\\nonumber
&= \frac{-1}{2\kappa} \int  \ud^2 S_j(\dot{h}_{ij} - \partial_j h_{0i} - \delta_{ij}\dot{h}_{kk}+ 2\delta_{ij}\partial_k h_{0k}- \partial_i h_{0j})
\\\nonumber
&= \frac{-1}{\kappa} \int  \ud^2 S_j(\Gamma^{(1)0}{}_{ij} - \delta_{ij} \Gamma^{(1)0}{}_{kk})
\\
&= \frac{-1}{\kappa} \int  \ud^2 S_j\pi^{(1)}_{ij},
\end{align}
which is the familiar expression for the ADM momentum \cite{ADMIV,ADMSummary} truncated at linear order. Thus, when the terms $O(h^3)$ can be neglected from the field equations, and the terms $O(h^2)$ can be neglected at spatial infinity, our gravitational Belinfante tensor $t_{\mu\nu}$ defines exactly the same total energy and momentum as ADM. Moreover,  these results would also arise if we had defined $\mathfrak{P}_{\mu}$ using $\tau_{\mu\nu}$, rather than the Belinfante tensor $t_{\mu\nu}$. Thus, $\tau_{\mu\nu}$ is able to cast the global information present in the ADM energy-momentum in terms of a local description with physically sensible properties, including gravitational energy-density that is nowhere negative and gravitational energy-flux that is nowhere spacelike.

Although the ADM energy-momentum is usually represented in terms of the asymptotic behaviour of the gravitational field, as above, the reader should also be aware that these global quantities can be cast as spatial integrals of a gravitational Belinfante tensor $t^\mathrm{ADM}_{\mu\nu}$ that emerges from the canonical formalism \cite{ADMIII,ADMIV,ADMSummary}. Although ADM did not propose that this tensor should be interpreted as a physically meaningful local measure of gravitational energy-momentum, it is nonetheless interesting to compare the quadratic part of $t^\mathrm{ADM}_{\mu\nu}$ with $\tau_{\mu\nu}$ and $t_{\mu\nu}$. The strongest resemblance occurs when we employ our gauge-fixing procedure (that is, we insist that $h_{\mu\nu}$ be transverse-traceless) and examine the $(0,0)$ components of the tensors:
\begin{align}\nonumber
\tau_{00}= t_{00}& =\frac{1}{8\kappa}(\dot{h}_{ij} \dot{h}_{ij} + \partial_k h_{ij}  \partial_k h_{ij})\\\label{thanks} &=  t^{\mathrm{ADM}}_{00} + O(h^3).
\end{align}
Remarkably, we find that our gauge-fixed $\tau_{00}$ and $t_{00}$ are in fact equal to the ADM  ``Hamiltonian density'' $t^{\mathrm{ADM}}_{00}$ when working to quadratic order. This is a rather surprising correspondence, particularly when one considers how little our framework has in common with the canonical 3+1 approach from which $t^{\mathrm{ADM}}_{00}$ arose. Note, however, that this equality does not extend to the other components of the tensors:
\begin{subequations}\label{manyt}
\begin{align}
\kappa\tau_{i0}&= \frac{1}{4}\dot{h}_{jk}\partial_i h_{jk}\\
\kappa t_{i0}&= \frac{1}{4}\left(\dot{h}_{jk}\partial_i h_{jk}-\dot{h}_{jk} \partial_j h_{ik} + h_{jk}\partial_j \dot{h}_{ki}\right)\\
\kappa t^{\mathrm{ADM}}_{i0}&=\frac{1}{4}\left(\dot{h}_{jk}\partial_i h_{jk}-2\dot{h}_{jk} \partial_j h_{ik}\right) +O(h^3).
\end{align}
\end{subequations}
Hence, $\tau_{\mu\nu}$, $t_{\mu\nu}$, and $t^{\mathrm{ADM}}_{\mu\nu}$ do not give rise to equivalent local descriptions of gravitational energy-momentum, even in transverse-traceless gauge; moreover, even though $t^{\mathrm{ADM}}_{\mu\nu}$ succeeds in defining a positive gravitational energy-density $t^{\mathrm{ADM}}_{00}\ge 0$ as seen by observers at rest with respect to the \TT-frame, it does not display the full (Lorentz-invariant) positivity properties of $\tau_{\mu\nu}$: the energy-density $v^\mu v^\nu t^{\mathrm{ADM}}_{\mu\nu}$ (as seen by an observer moving with 4-velocity $v^\mu$) may be negative, and the energy-flux $v^\mu t^{\mathrm{ADM}}_{\mu\nu}$ may be spacelike.

The formulae (\ref{manyt}) also serve as another starting point from which to verify that all three tensors define the same total momentum $\mathfrak{P}_i$: those terms in $t_{i0}$ and $t^{\mathrm{ADM}}_{i0}$ which do not appear in $\tau_{i0}$ can be integrated by parts (discarding a quadratic surface term, as usual) and then vanish due to the gauge condition $\partial_i h_{ij}=0$. Of course, these terms \emph{do} contribute to the total \emph{angular momentum}: they correspond to the divergence $\partial_\alpha (s_{\mu\nu}^{\phantom{\mu\nu}\alpha} + s_{\nu\mu}^{\phantom{\mu\nu}\alpha} - s^\alpha_{\phantom{\alpha}\mu\nu})/2$ that packages intrinsic spin into the Belinfante tensor (\ref{Beldef}). Accordingly, when one neglects quadratic surface terms, one finds that
\begin{align}\nonumber
\int (2x_{[i}\tau_{j]}{}^0 + s^0_{\phantom{0}ij})\ud^3 x & = \int2x_{[i}t_{j]}{}^{0} \ud^3 x \\
&= \int 2x_{[i}t^{\mathrm{ADM}}{}_{j]}{}^{0}\ud^3 x + O(h^3),
\end{align}
confirming that $\tau_{\mu\nu}$ and $s^{\alpha}_{\phantom{\alpha}\mu\nu}$ also give the same global description of angular momentum as ADM at second order.

In summary, whenever general relativity can be approximated to quadratic order, and quadratic surface terms can be neglected at spatial infinity, our gravitational energy-momentum tensor $\tau_{\mu\nu}$ and spin tensor $s^{\alpha}_{\phantom{\alpha}\mu\nu}$ provide the same \emph{global} description of energy, momentum, and angular momentum as ADM, but localise these quantities in a physically sensible fashion, displaying positive gravitational energy-density, causal energy-flux, and spatial spin.

\bibliography{TD}
\end{document}